\documentclass[aps,showpacs,preprintnumbers,amsmath, amssymb]{revtex4}

\usepackage{float}
\usepackage{graphics,epsfig}
\usepackage{graphicx}
\usepackage{dcolumn}
\usepackage{bm}

\begin{document}
\baselineskip=0.8 cm
\title{{\bf Backreacting holographic superconductors from the coupling of a scalar field to the Einstein tensor}}

\author{Dong Wang$^{1}$, Manman Sun$^{1}$, Qiyuan Pan$^{1,2}$\footnote{panqiyuan@126.com}, and Jiliang Jing$^{1,2}$\footnote{jljing@hunnu.edu.cn}}
\affiliation{$^{1}$Key Laboratory of Low Dimensional Quantum
Structures and Quantum Control of Ministry of Education, Synergetic
Innovation Center for Quantum Effects and Applications, and
Department of Physics, Hunan Normal University, Changsha, Hunan
410081, China} \affiliation{$^{2}$Center for Gravitation and
Cosmology, College of Physical Science and Technology, Yangzhou
University, Yangzhou 225009, China}

\vspace*{0.2cm}
\begin{abstract}
\baselineskip=0.6 cm
\begin{center}
{\bf Abstract}
\end{center}

We investigate the properties of the backreacting holographic
superconductors from the coupling of a scalar field to the Einstein
tensor in the background of a $d$-dimensional AdS black hole.
Imposing the Dirichlet boundary condition of the trial function
without the Neumann boundary conditions, we improve the analytical
Sturm-Liouville method with an iterative procedure to explore the
pure effect of the Einstein tensor on the holographic
superconductors and find that the Einstein tensor hinders the
condensate of the scalar field but does not affect the critical
phenomena. Our analytical findings are in very good agreement with
the numerical results from the ``marginally stable modes" method,
which implies that the Sturm-Liouville method is still powerful to
study the holographic superconductors from the coupling of a scalar
field to the Einstein tensor even if we consider the backreactions.

\end{abstract}

\maketitle
\newpage
\vspace*{0.2cm}

\section{Introduction}

It is well known that the superconductivity is one of the most
remarkable phenomena observed in physics in the 20th century
\cite{Tinkham}. However, the core mechanism of the high-temperature
superconductor systems, which can not be described by the usual
Bardeen-Cooper-Schrieffer (BCS) theory \cite{BCS}, is still one of
the unsolved mysteries in theoretical physics so far. Interestingly,
it was suggested that it is logical to investigate the properties of
high temperature superconductors on the boundary of spacetime by
considering the classical general relativity in one higher dimension
with the help of the Anti-de Sitter/conformal field theories
(AdS/CFT) correspondence \cite{Maldacena,Gubser1998,Witten}. In the
probe limit, Gubser observed that the spontaneous $U(1)$ symmetry
breaking by bulk black holes can be used to construct gravitational
dual of the transition from normal state to superconducting state
\cite{GubserPRD78}, and Hartnoll \emph{et al.} found that the
properties of a ($2+1$)-dimensional superconductor can indeed be
reproduced in the ($3+1$)-dimensional holographic dual model based
on the framework of usual Maxwell electrodynamics
\cite{HartnollPRL101}. Extending the investigation to the so-called
holographic superconductor models away from the probe limit, i.e.,
taking the backreactions of the spacetime into account, the authors
of Ref. \cite{HartnollJHEP12} showed that even the uncharged scalar
field can form a condensate in the $(2+1)$-dimensional holographic
superconductor model. Along this line, there has been accumulated
interest in studying the effects of the backreaction on the
holographic $s$-wave
\cite{PanJWCh,BarclayGregory,Barclay2011,GregoryRev,KannoCQG,Ge2011,Herzog2010,Gubser-Nellore,
Horowitz-Way,Aprile-Russo,Brihaye,Liu-Sun,Siani,PanWangBR,LiuWangBTZ,LPW2012,GangopadhyayPLB,
LiuGongWang,YaoJing,EmparanTanabe,DeyJHEP2014,NakoniecznyRogatko,MomeniPLB2015,Ghorai2016,
JingCQG,SheykhiShakerPLB,Peng2017,SheykhiShakerIJMPD,SherkatghanadIJMPD,YaoJing2018,Ghotbabadi2018,GhoraiNPB2018,MohammadiSZ2018}
$p$-wave
\cite{CaiNieZhang2011,AriasLandea,CaiPWave-1,CaiPWave-2,CSJHEP2015,NieCai2015,Wang2016,Nie2017}
and $d$-wave \cite{GTWjhep2012} dual models. Reviews of the
holographic superconductors can be found in Refs.
\cite{HartnollRev,HerzogRev,HorowitzRev,CaiRev}.

Most of the aforementioned works on the gravitational dual models
focus on the superconductors without an impurity. As a matter of
fact, to study the effect of impurities is often important since
their presence can drastically change the physical properties of the
superconductors in condensed matter physics \cite{Balatsky}.
According to the AdS/CFT duality, Ishii and Sin investigated the
impurity effect in a holographic superconductor by turning on a
coupling between the gauge field and a new massive gauge field, and
found that the mass gap in the optical conductivity disappears when
the coupling is sufficiently large \cite{Ishii}. Zeng and Zhang
studied the single normal impurity effect in a superconductor by
using the holographic approach, which showed that the critical
temperature of the host superconductor decreases as the size of the
impurity increases and the phase transition at the critical impurity
strength (or the critical temperature) is of zeroth order
\cite{ZengZhang}. Fang \emph{et al.} extended the study to the
Fermionic phase transition induced by the effective impurity in
holography and obtained a phase diagram in $(\alpha,T)$ plane
separating the Fermi liquid phase and the non-Fermi liquid phase
\cite{FangJHEP}. More recently, Kuang and Papantonopoulos built a
holographic superconductor with a scalar field coupled kinematically
to the Einstein tensor and observed that, as the strength of the
coupling increases, the critical temperature below which the scalar
field condenses is lowering, the condensation gap decreases faster
than the temperature, the width of the condensation gap is not
proportional to the size of the condensate and at low temperatures
the condensation gap tends to zero for the strong coupling
\cite{KuangE2016}. Obviously, these effects suggest that the
derivative coupling in the gravity bulk can have a dual
interpretation on the boundary corresponding to impurities
concentrations in a real material. Note that they concentrated on
the probe limit where the backreaction of matter fields on the
spacetime metric is neglected. Thus, in this work we will extend
their interesting model to the case away from the probe limit and
explore the effect of the Einstein tensor on the holographic
superconductors with backreactions. In addition, we will compare the
result in five dimensions with that in four dimensions and present
an analysis of the effect the extra dimension has on the scalar
condensation formation. In the calculation, we first use the
Sturm-Liouville eigenvalue problem \cite{Siopsis,SiopsisBF} to
analytically study the holographic superconductor phase transition,
and then count on the ``marginally stable modes" method
\cite{GubserPRD78,marginally stable modes} to numerically confirm
the analytical findings and verify the effectiveness of the
Sturm-Liouville method.

The organization of the work is as follows. In Sec. II, we will
introduce the backreacting holographic superconductor models from
the coupling of a scalar field to the Einstein tensor in the
$d$-dimensional AdS black hole background. In Sec. III we will give
an analytical investigation of the holographic superconductors by
using the Sturm-Liouville method. In Sec. IV we will give a
numerical investigation of the holographic superconductors by using
the ``marginally stable modes" method. We will summarize our results
in the last section.

\section{Description of the holographic dual system}

The general action describing a charged, complex scalar field
coupled to the Einstein tensor $G^{\mu\nu}$ in the $d$-dimensional
Einstein-Maxwell action with negative cosmological constant
$\Lambda=-(d-1)(d-2)/(2L^{2})$ is of the form
\begin{eqnarray}\label{System}
S=\int
d^{d}x\sqrt{-g}\left[\frac{1}{2\kappa^{2}}\left(R-2\Lambda\right)
-\frac{1}{4}F_{\mu\nu}F^{\mu\nu}-(g^{\mu\nu}+\eta G^{\mu\nu})D_{\mu}\psi(D_{\nu}\psi)^{\ast}
-m^2|\psi|^2\right],
\end{eqnarray}
with $D_{\mu}=\nabla_{\mu}-iqA_{\mu}$. Here $\kappa^{2}=8\pi G_{d}$
represents the gravitational constant,
$F_{\mu\nu}=\nabla_{\mu}A_{\nu}-\nabla_{\nu}A_{\mu}$ is the Maxwell
field strength, and $\psi$ denotes the scalar field with the charge
$q$ and mass $m$. When the coupling parameter $\eta\rightarrow0$,
our model reduces to the standard holographic superconductors with
backreactions investigated in \cite{HartnollJHEP12,PanJWCh}. It
should be noted that we can rescale the bulk fields $\psi$ and
$A_{\mu}$ as $\psi/q$ and $A_{\mu}/q$ in order to put the factor
$1/q^{2}$ as the backreaction parameter for the matter fields. So
the probe limit can be obtained safely if
$\kappa^{2}/q^{2}\rightarrow0$. Without loss of generality, we can
set $q=1$ and keep $\kappa^{2}$ finite when we take the backreaction
into account, just as in Refs.
\cite{PanJWCh,BarclayGregory,Barclay2011,GregoryRev,KannoCQG,CSJHEP2015}.

To go beyond the probe limit, we adopt the metric ansatz for the
black hole with the curvature $k=0$ as
\begin{eqnarray}\label{BH metric}
ds^2=-f(r)e^{-\chi(r)}dt^{2}+\frac{dr^2}{f(r)}+r^{2}h_{ij}dx^{i}dx^{j},
\end{eqnarray}
where $f$ and $\chi$ are functions of $r$ only, $h_{ij}dx^{i}dx^{j}$
represents the line element of a ($d-2$)-dimensional hypersurface.
Obviously, the Hawking temperature of this $d$-dimensional black
hole, which will be interpreted as the temperature of the CFT, can
be given by
\begin{eqnarray}\label{Hawking temperature}
T_{H}=\frac{f^{\prime}(r_{+})e^{-\chi(r_{+})/2}}{4\pi},
\end{eqnarray}
where the prime denotes a derivative with respect to $r$, and the
black hole horizon $r_{+}$ is determined by $f(r_{+})=0$. For the
considered ansatz (\ref{BH metric}), the nonzero components of the
Einstein tensor $G^{\mu\nu}$ are
\begin{eqnarray}\label{Einstein tensor}
&&G^{tt}=-\frac{(d-2)e^{\chi}}{2r^{2}}\left[(d-3)+\frac{rf^{\prime}}{f}\right],~~
G^{rr}=\frac{(d-2)f^{2}}{2r^{2}}\left[(d-3)+\frac{rf^{\prime}}{f}-r\chi^\prime\right],\nonumber \\
&&G^{xx}=G^{yy}=\cdot\cdot\cdot=\frac{1}{4r^{3}}\left\{f^\prime\left[4(d-3)-3r\chi^\prime\right]+2rf''+f\left[\frac{2(d-3)(d-4)}{r}-2(d-3)\chi^\prime+r\chi'^{2}-2r\chi''\right]\right\}.
\nonumber \\
\end{eqnarray}
For the scalar field and electromagnetic field, we will take
$\psi=|\psi|$, $A_{t}=\phi$ where $\psi$, $\phi$ are both real
functions of $r$ only. Thus, from the action (\ref{System}) we can
give the equations of motion for the metric functions $f(r)$ and
$\chi(r)$
\begin{eqnarray}\label{chi}
\bigg[1+\eta\kappa^{2}\bigg(\frac{e^{\chi}\phi^{2}\psi^{2}}{f}-3f\psi'^{2}\bigg)\bigg]\chi^{\prime}&+&\nonumber\\\frac{4\kappa^{2}r}{d-2}\bigg\{\psi'^{2}+\frac{e^{\chi}\phi^{2}\psi^{2}}{f^{2}}
+\frac{(d-2)\eta}{2r}\bigg[\frac{(d-3)f}{r}\left(\psi'^{2}+\frac{e^{\chi}\phi^{2}\psi^{2}}{f^{2}}\right)+\frac{2e^{\chi}\phi\psi(\phi\psi)'}{f}-2f\psi'\psi''\bigg]\bigg\}&=&0,
\end{eqnarray}
\begin{eqnarray}\label{fr}
\bigg[1+\eta\kappa^{2}\left(\frac{e^{\chi}\phi^{2}\psi^{2}}{f}-3f\psi'^{2}\right)\bigg]f^{\prime}-\left[\frac{(d-1)r}{L^{2}}-\frac{(d-3)f}{r}\right]+\frac{2\kappa^{2}r}{d-2}\bigg\{m^{2}\psi^{2}+\frac{e^{\chi}\phi'^{2}}{2}&+&\nonumber\\
f\bigg(\psi'^{2}+\frac{e^{\chi}\phi^{2}\psi^{2}}{f^{2}}\bigg)+\frac{(d-2)\eta}{2r}\bigg[\frac{(d-3)e^{\chi}\phi^{2}\psi^{2}}{r}-\frac{(d-3)f^{2}\psi'^{2}}{r}-4f^{2}\psi'\psi''\bigg]\bigg\}&=&0,
\end{eqnarray}
and for the matter fields $\phi(r)$ and $\psi(r)$
\begin{eqnarray}
\phi^{\prime\prime}+\left(\frac{d-2}{r}+\frac{\chi^{\prime}}{2}\right)\phi^\prime-\frac{2\psi^{2}}{f}\left[1+\frac{(d-2)\eta
f}{2r}\bigg(\frac{d-3}{r}+\frac{f'}{f}\bigg)\right]\phi=0,\label{phi}
\end{eqnarray}
\begin{eqnarray}
\left\{1+\frac{(d-2)\eta}{2r}\bigg[\frac{(d-3)f}{r}+f'-f\chi'\bigg]\right\}\psi^{\prime\prime}+\bigg\{\left(\frac{d-2}{r}+\frac{f'}{f}-\frac{\chi'}{2}\right)&+&\nonumber\\
\frac{(d-2)\eta}{2r}\bigg[f''+\frac{3(d-3)f'}{r}+\frac{f'^{2}}{f}+\frac{f\chi'^{2}}{2}-\frac{3(d-3)f\chi'}{2r}-\frac{5f'\chi'}{2}-f\chi''+\frac{(d-3)(d-4)f}{r^{2}}\bigg]\bigg\}\psi^\prime
&+&\nonumber \\
\left\{\frac{e^{\chi}\phi^{2}}{f^{2}}\bigg[1+\frac{(d-2)(d-3)\eta
f}{2r^{2}}+\frac{(d-2)\eta
f'}{2r}\bigg]-\frac{m^{2}}{f}\right\}\psi&=&0,\label{psi}
\end{eqnarray}
where the prime denotes a derivative with respect to $r$.

We will count on the appropriate boundary conditions to get the
solutions in the superconducting phase, $\psi(r)\neq0$. At the
horizon $r=r_{+}$ of the black hole, the regularity gives the
boundary conditions
\begin{eqnarray}\label{HorizonBC}
&&\phi(r_{+})=0\,,\hspace{0.5cm}
\psi(r_{+})=\frac{1}{m^{2}}\left[f^\prime(r_{+})+\frac{(d-2)\eta f^\prime(r_{+})^{2}}{2r_{+}}\right]\psi^\prime(r_{+}),\nonumber\\
&&\chi^{\prime}(r_{+})=-\frac{4\kappa^{2}r_{+}}{d-2}\bigg[\psi'(r_{+})^{2}+\frac{e^{\chi(r_{+})}\phi'(r_{+})^{2}\psi(r_{+})^{2}}{f'(r_{+})^{2}}+\frac{(d-2)\eta e^{\chi(r_{+})}\phi'(r_{+})^{2}\psi(r_{+})^{2}}{r_{+}f'(r_{+})}\bigg],\nonumber\\
&&f^{\prime}(r_{+})=\frac{(d-1)r_{+}}{L^{2}}-\frac{2\kappa^{2}
r_{+}}{d-2}\left[m^{2}\psi(r_{+})^{2}+\frac{1}{2}e^{\chi(r_{+})}\phi^{\prime}(r_{+})^{2}\right].
\end{eqnarray}
At the asymptotic boundary $r\rightarrow\infty$, the solutions behave like
\begin{eqnarray}
\chi\rightarrow0\,,\hspace{0.5cm}
f\sim\frac{r^{2}}{L^{2}}\,,\hspace{0.5cm}
\phi\sim\mu-\frac{\rho}{r^{d-3}}\,,\hspace{0.5cm}
\psi\sim\frac{\psi_{-}}{r^{\Delta_{-}}}+\frac{\psi_{+}}{r^{\Delta_{+}}}\,,
\label{InfinityBC}
\end{eqnarray}
with the characteristic exponent
\begin{eqnarray}
\Delta_\pm=\frac{1}{2}\left[(d-1)\pm\sqrt{(d-1)^{2}+\frac{8m^{2}L^{2}}{2+(d-1)(d-2)\eta}}~\right].
\label{CharacteristicExponent}
\end{eqnarray}
According to the AdS/CFT correspondence, $\mu$ and $\rho$ are
interpreted as the chemical potential and charge density in the dual
field theory, respectively. Considering the stability of the scalar
field, we find that the mass should be above the
Breitenlohner-Freedman (BF) bound
$m^{2}_{BF}=-(d-1)^{2}[2+(d-1)(d-2)\eta]/(8L^{2})$
\cite{Breitenloher}, which depends on the coupling parameter $\eta$
and dimensionality of the AdS space $d$. Note that, provided
$\Delta_{-}$ is larger than the unitarity bound, both $\psi_{-}$ and
$\psi_{+}$ can be normalizable and be used to define operators on
the dual field theory, $\psi_{-}=\langle{\cal O}_{-}\rangle$,
$\psi_{+}=\langle{\cal O}_{+}\rangle$, respectively
\cite{HartnollPRL101,HartnollJHEP12}.  In this work, we impose
boundary condition $\psi_{-}=0$ since we concentrate on the
condensate for the operator $\langle{\cal O}_{+}\rangle$.

In the normal phase, $\psi(r)=0$, the metric coefficient $\chi$ is a
constant from Eq. (\ref{chi}) and the analytical solutions to Eqs.
(\ref{fr}) and (\ref{phi}) lead to the Reissner-Nordstr\"{o}m AdS
black holes
\begin{eqnarray}\label{RNBH}
f=\frac{r^{2}}{L^{2}}-\frac{1}{r^{d-3}}\left[\frac{r^{d-1}_{+}}{L^{2}}+\frac{(d-3)\kappa^{2}\rho^{2}}{(d-2)r^{d-3}_{+}}\right]
+\frac{(d-3)\kappa^{2}\rho^{2}}{(d-2)r^{2(d-3)}}\,,\hspace{0.5cm}
\phi=\mu-\frac{\rho}{r^{d-3}},
\end{eqnarray}
where the metric coefficient $f$ goes back to the case of the
Schwarzschild AdS black hole when $\kappa=0$.

On the other hand, from the equations of motion (\ref{chi})-(\ref{psi}) we can obtain
the useful scaling symmetries and the transformation of the relevant quantities
\begin{eqnarray}
&&r\rightarrow\alpha r,~~(t,x^{i})\rightarrow\frac{1}{\alpha}(t,x^{i}),\nonumber\\
&&(\chi,\psi,L)\rightarrow(\chi,\psi,L),~~(\phi,\mu,T)\rightarrow\alpha(\phi,\mu,T),\nonumber\\
&&\rho\rightarrow\alpha^{d-2}\rho,~~\psi_{\pm}\rightarrow\alpha^{\Delta_{\pm}}\psi_{\pm},
\end{eqnarray}
where $\alpha$ is a real positive number. For simplicity, we use the scaling symmetries to set $L=1$ when
performing calculations in the following sections.

\section{Critical behavior from the Sturm-Liouville method}

We use the variational method for the Sturm-Liouville eigenvalue
problem \cite{Siopsis,SiopsisBF} to analytically investigate the
properties of the backreacting holographic superconductors from the
coupling of a scalar field to the Einstein tensor. We will derive
the critical behavior of the system near the phase transition point
and examine the effects of the Einstein tensor and backreaction on
the holographic superconductors.

For convenience, we introduce a new variable $z=r_{+}/r$ and rewrite the equations of motion (\ref{chi})-(\ref{psi}) into
\begin{eqnarray}
\bigg[1+\eta\kappa^{2}\bigg(\frac{e^{\chi}\phi^{2}\psi^{2}}{f}-\frac{3z^{4}f\psi'^{2}}{r_{+}^{2}}\bigg)\bigg]\chi^{\prime}-\frac{4\kappa^{2}}{d-2}\bigg\{z\psi'^{2}+\frac{r_{+}^{2}e^{\chi}\phi^{2}\psi^{2}}{z^{3}f^{2}}&+&\nonumber \\
\frac{(d-2)\eta}{2}\bigg[\frac{(d-3)e^{\chi}\phi^{2}\psi^{2}}{zf}-\frac{2e^{\chi}\phi\psi(\phi\psi)^\prime}{f}+\frac{(d+1)z^{3}f\psi'^{2}}{r_{+}^{2}}+\frac{2z^{4}f\psi'\psi''}{r_{+}^{2}}\bigg]\bigg\}&=&0,\label{chiz}
\end{eqnarray}
\begin{eqnarray}\label{frz}
\bigg[1+\eta\kappa^{2}\bigg(\frac{\phi^{2}\psi^{2}e^{\chi}}{f}-\frac{3z^{4}f\psi'^{2}}{r_{+}^{2}}\bigg)\bigg]f^{\prime}-\frac{(d-3)f}{z}+\frac{(d-1)r^{2}_{+}}{z^{3}}-\frac{2r_{+}^{2}\kappa^{2}}{(d-2)z^{3}}\bigg\{m^{2}\psi^{2}+\frac{e^{\chi}z^{4}\phi'^{2}}{2r_{+}^{2}}&+&\nonumber \\
f\bigg(\frac{e^{\chi}\phi^{2}\psi^{2}}{f^{2}}+\frac{z^{4}\psi'^{2}}{r_{+}^{2}}\bigg)+\frac{(d-2)\eta}{2r_{+}^{2}}\bigg[(d-3)z^{2}\phi^{2}\psi^{2}e^{\chi}-\frac{(d-11)z^{6}f^{2}\psi'^{2}}{r_{+}^{2}}+\frac{4z^{7}f^{2}\psi'\psi''}{r_{+}^{2}}\bigg]\bigg\}&=&0,
\end{eqnarray}
\begin{eqnarray}\label{phiz}
\phi^{\prime\prime}+\left(\frac{\chi'}{2}-\frac{d-4}{z}\right)\phi^{\prime}-\frac{2r_{+}^{2}\psi^{2}}{z^{4}f}\bigg[1+\frac{(d-2)\eta
z^{2}f}{2r_{+}^{2}}\bigg(d-3-\frac{zf'}{f}\bigg)\bigg]\phi=0,
\end{eqnarray}
\begin{eqnarray}\label{psiz}
\bigg[1+\frac{(d-2)\eta z^{2}f}{2r_{+}^{2}}\bigg(d-3+z\chi'-\frac{zf'}{f}\bigg)\bigg]\psi^{\prime\prime}-\bigg\{\frac{\chi'}{2}+\frac{d-4}{z}-\frac{f'}{f}+\frac{(d-2)\eta}{2r_{+}^{2}}\bigg[(d-3)(d-6)zf&+&\nonumber\\
\frac{z^{3}f'^{2}}{f}-(3d-13)z^{2}f'+z^{3}f''+\frac{z^{3}f\chi'^{2}}{2}-\frac{5z^{3}f'\chi'}{2}+\frac{(3d-17)z^{2}f\chi'}{2}-z^{3}f\chi''\bigg]\bigg\}\psi^\prime&+&\nonumber \\
\frac{r_{+}^{2}e^{\chi}\phi^{2}}{z^{4}f^{2}}\bigg[1+\frac{(d-2)\eta
z^{2}f}{2r_{+}^{2}}\bigg(d-3-\frac{zf'}{f}\bigg)\bigg]\psi-\frac{m^{2}r_{+}^{2}}{z^{4}f}\psi&=&0,
\end{eqnarray}
where the prime now denotes the derivative with respect to $z$. Note
that the scalar field $\psi=0$ at the critical temperature $T_{c}$.
Therefore the expectation value of the scalar operator $\langle{\cal
O}_{+}\rangle$ is small near the critical point and we can select it
as an expansion parameter $\epsilon\equiv\langle{\cal O}_{+}\rangle$
with $\epsilon\ll1$. Since we are interested in solutions where
$\psi$ is small, so from Eqs. (\ref{phiz}) and (\ref{psiz}) we can
expand the scalar field $\psi(z)$ and the gauge field $\phi(z)$ as
\cite{PanJWCh,KannoCQG,Ge2011,Herzog2010}
\begin{eqnarray}
&&\psi=\epsilon\psi_{1}+\epsilon^{3}\psi_{3}+\epsilon^{5}\psi_{5}+\cdot\cdot\cdot,\nonumber\\
&&\phi=\phi_{0}+\epsilon^{2}\phi_{2}+\epsilon^{4}\phi_{4}+\cdot\cdot\cdot,
\end{eqnarray}
and from Eqs. (\ref{chiz}) and (\ref{frz}) the metric functions
$f(z)$ and $\chi(z)$ can be expanded around the
Reissner-Nordstr\"{o}m AdS spacetime as
\begin{eqnarray}
&&f=f_{0}+\epsilon^{2}f_{2}+\epsilon^{4}f_{4}+\cdot\cdot\cdot,\nonumber\\
&&\chi=\epsilon^{2}\chi_{2}+\epsilon^{4}\chi_{4}+\cdot\cdot\cdot.
\end{eqnarray}
Considering that the chemical potential $\mu$ can be corrected
order by order $\mu=\mu_{0}+\epsilon^{2}\delta\mu_{2}+\cdot\cdot\cdot$ with $\delta\mu_{2}>0$ \cite{Herzog2010},
we get a result for the order parameter as a function of the chemical
potential near the phase transition
\begin{eqnarray}
\epsilon\equiv\langle{\cal
O}_{+}\rangle\approx\left(\frac{\mu-\mu_{0}}{\delta\mu_{2}}\right)^{1/2},
\end{eqnarray}
which indicates that the holographic s-wave superconducting phase
transition with backreaction from the coupling of a scalar field to
the Einstein tensor is of the second order and the critical exponent
of the system always takes the mean-field value 1/2. The Einstein
tensor, backreaction and spacetime dimension will not influence the
result. When $\mu\rightarrow\mu_{0}$, the phase transition occurs
and the order parameter is zero at the critical point, which means
that the critical value $\mu$ is $\mu_{c}=\mu_{0}$.

Now we are in a position to solve equations order by order. At the
zeroth order, the equation of motion for the Maxwell field
(\ref{phiz}) reduces to
\begin{eqnarray}
\phi_{0}^{\prime\prime}(z)-\frac{d-4}{z}\phi_{0}^\prime(z)=0,\label{phi0}
\end{eqnarray}
which has a solution $\phi_{0}(z)=\mu_{0}(1-z^{d-3})$ with
$\mu_{0}=\rho/r^{d-3}_{+}$. Since
$\mu_{0}=\mu_{c}=\rho/r^{d-3}_{+c}$ at the critical point $\mu_{c}$,
where $r_{+c}$ is the radius of the horizon at the critical point,
we will have
\begin{eqnarray}
\phi_{0}(z)=\lambda r_{+c}(1-z^{d-3}), \label{Phi-critical}
\end{eqnarray}
where we have set a dimensionless quantity
$\lambda=\rho/r^{d-2}_{+c}$. Inserting this solution into Eq.
(\ref{frz}), we can give the equation of motion for the metric
function $f_{0}(z)$, i.e.,
\begin{eqnarray}\label{f0z}
f_{0}^{\prime}-\frac{(d-3)f_{0}}{z}+\frac{(d-1)r^{2}_{+}}{z^{3}}-\frac{\kappa^{2}z\phi_{0}'^{2}}{d-2}=0,
\end{eqnarray}
with its solution
\begin{eqnarray}
f_{0}(z)=r^{2}_{+}\xi(z)=r^{2}_{+}\left[\frac{1}{z^{2}}-z^{d-3}-\frac{(d-3)\kappa^{2}\lambda^{2}}{d-2}z^{d-3}(1-z^{d-3})\right],
\label{f-critical}
\end{eqnarray}
where we have defined a new function $\xi(z)$ for simplicity.

\begin{figure}[ht]
\includegraphics[scale=0.63]{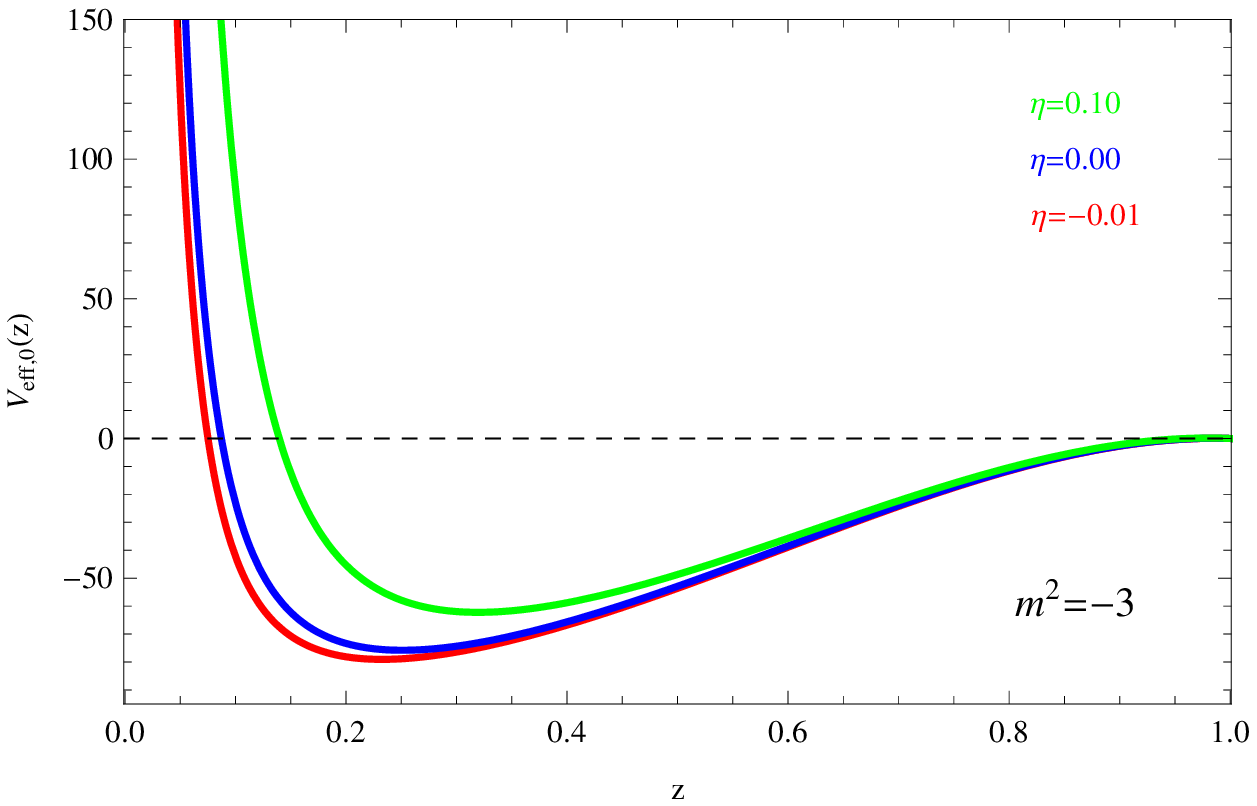}\hspace{0.2cm}%
\includegraphics[scale=0.63]{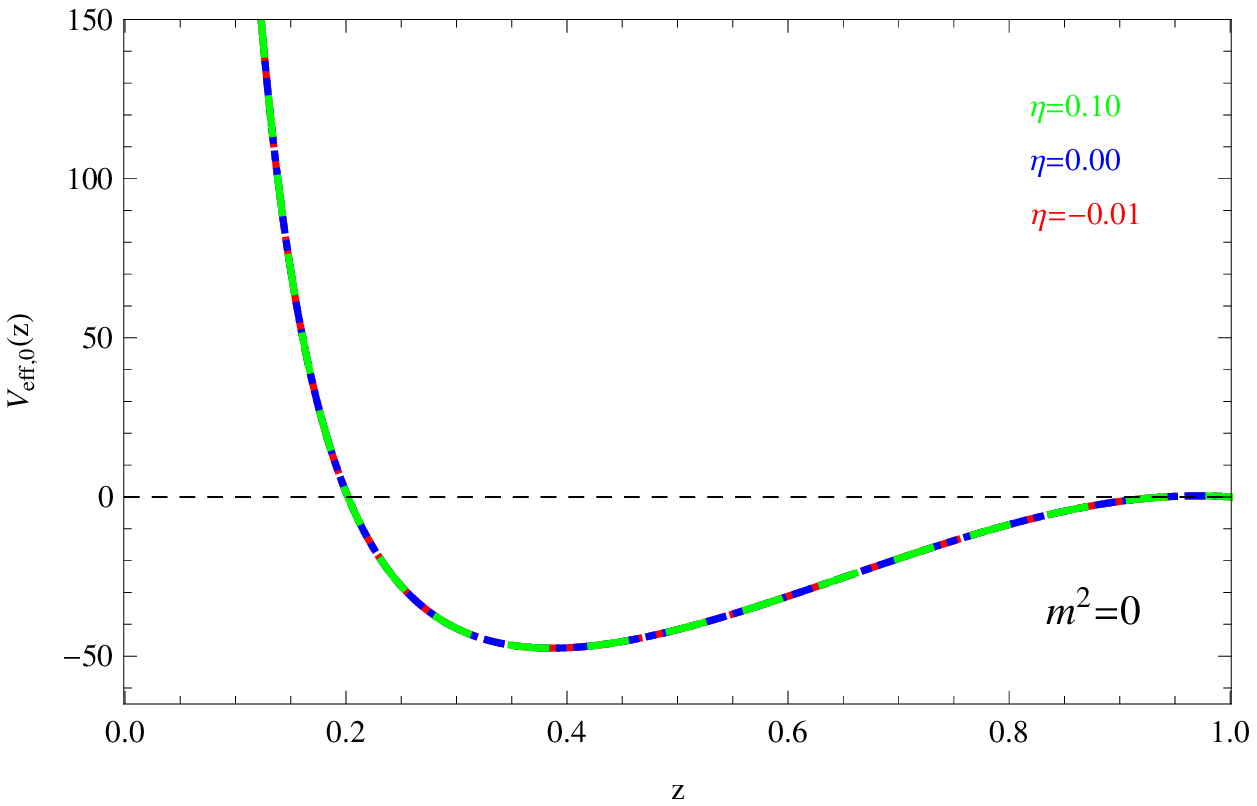}\\ \vspace{0.0cm}
\includegraphics[scale=0.63]{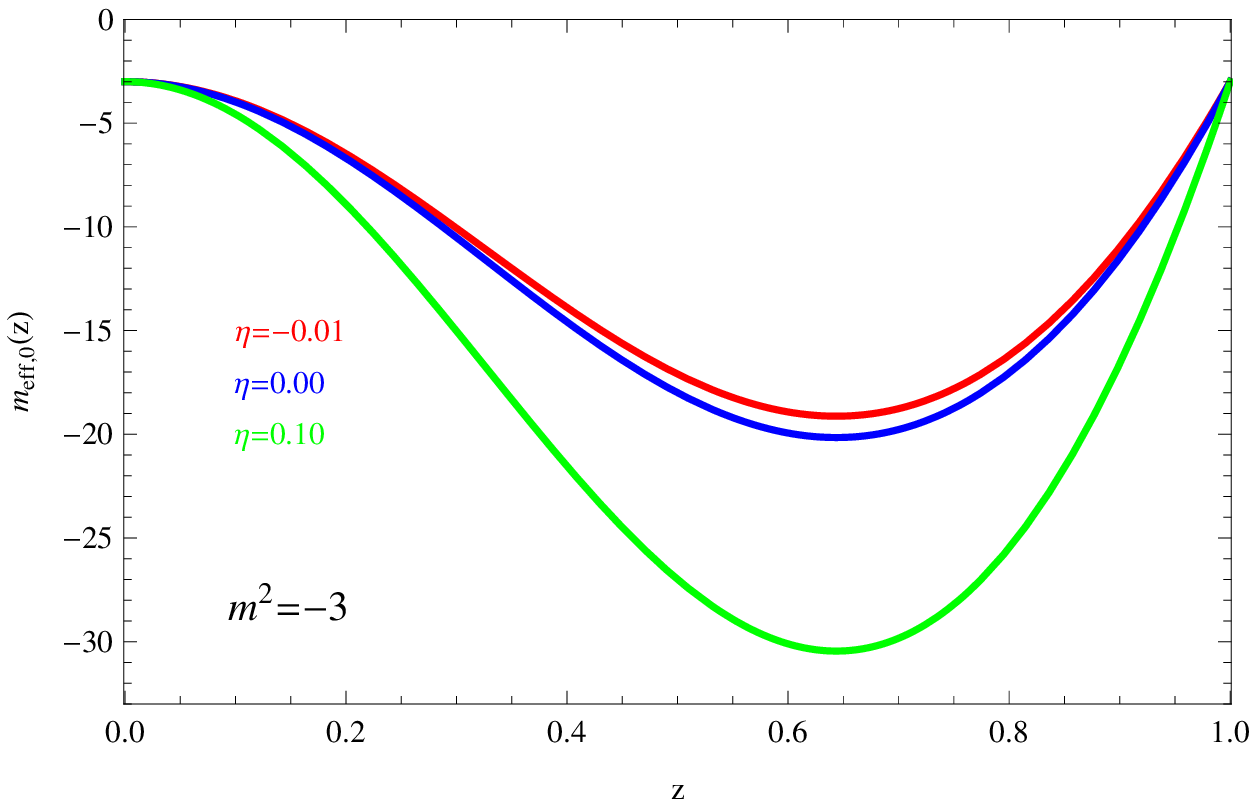}\hspace{0.2cm}%
\includegraphics[scale=0.63]{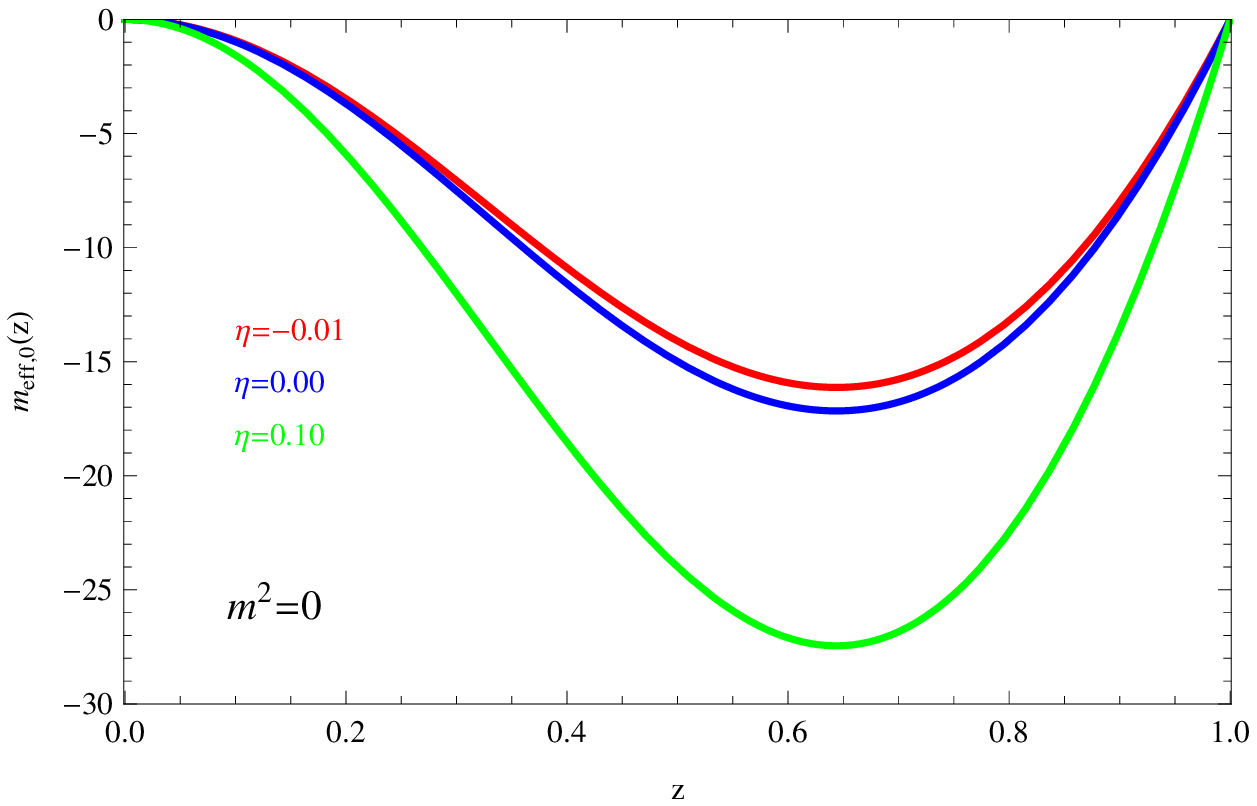}\\ \vspace{0.0cm}
\caption{\label{VeffMeff} (Color online) The effective potential
$V_{eff,0}(z)$ and effective mass $m_{eff,0}(z)$ as a function of
the radial coordinate $z$ outside the horizon with the coupling
parameter $\eta=-0.01$ (red), 0.00 (blue), and 0.10 (green) for the
fixed mass of the scalar field $m^{2}=-3$ (left) and $m^{2}=0$
(right), backreaction parameter $\kappa=0$ and dimensionless
quantity $\lambda=10$ in $d=5$ dimensions. It should be noted that
the three lines of the effective potential coincide in the top-right
panel. }
\end{figure}

At the first order, the equation of motion for $\psi_{1}(z)$ is
\begin{eqnarray}\label{psi1z}
\bigg[1+\frac{(d-2)\eta z^{2}f_{0}}{2r_{+}^{2}}\bigg(d-3-\frac{zf_{0}'}{f_{0}}\bigg)\bigg]\psi_{1}^{\prime\prime}-\bigg\{\frac{d-4}{z}-\frac{f_{0}'}{f_{0}}+\frac{(d-2)\eta}{2r_{+}^{2}}\bigg[(d-3)(d-6)zf_{0}+\frac{z^{3}f_{0}'^{2}}{f_{0}}&-&\nonumber\\
(3d-13)z^{2}f_{0}'+z^{3}f_{0}''\bigg]\bigg\}\psi_{1}^\prime+\frac{r_{+}^{2}\phi_{0}^{2}}{z^{4}f_{0}^{2}}\bigg[1+\frac{(d-2)\eta
z^{2}f_{0}}{2r_{+}^{2}}\bigg(d-3-\frac{zf_{0}'}{f_{0}}\bigg)\bigg]\psi_{1}-\frac{m^{2}r_{+}^{2}}{z^{4}f_{0}}\psi_{1}&=&0,
\end{eqnarray}
which has the asymptotic AdS boundary condition
$\psi_{1}\sim\frac{\psi_{-}}{r_{+}^{\Delta_{-}}}z^{\Delta_{-}}+\frac{\psi_{+}}{r_{+}^{\Delta_{+}}}z^{\Delta_{+}}$.
Just as in the interesting works by Kolyvaris \emph{et al.}
\cite{KolyvarisKPSA,KolyvarisKPSB}, we will use Eq. (25) to discuss
the stability of our solutions. We can express the effective
potential of $\psi_{1}$ as
\begin{eqnarray}
V_{eff,0}(z)=-\frac{f^{2}_{0}}{r_{+}^{4}}\left\{\frac{z^{4}P''(z)}{P(z)}-\frac{r_{+}^{d-2}W(z)P(z)P'(z)}{z^{d-5}}
-\frac{r_{+}^{d}P^{2}(z)}{z^{d-2}f_{0}}\left[m^{2}-\frac{z^{d-2}\phi^{2}_{0}}{r_{+}^{d-2}f_{0}P^{2}(z)}\right]\right\},
\end{eqnarray}
with the defined functions
\begin{eqnarray}
P(z)&=&\left(\frac{z}{r_{+}}\right)^\frac{d-2}{2}\bigg[1+\frac{(d-2)\eta
z^{2}f_{0}}{2r_{+}^{2}}\bigg(d-3-\frac{zf_{0}'}{f_{0}}\bigg)\bigg]^{-\frac{1}{2}},\nonumber\\
W(z)&=&(d-4)-\frac{zf'_{0}}{f_{0}}+\frac{(d-2)\eta
z^{2}}{2r_{+}^{2}}\bigg[(d-3)(d-6)f_{0}+\frac{z^{2}f_{0}'^{2}}{f_{0}}
-(3d-13)zf_{0}'+z^{2}f_{0}''\bigg],
\end{eqnarray}
which can develop a negative gap near the black hole horizon,
implying a potential instability of the black hole. In Fig.
\ref{VeffMeff}, we plot the curves of the effective potential
$V_{eff,0}(z)$ with different values of the coupling parameter
$\eta$ for the fixed mass of the scalar field $m^{2}=-3$ (top-left)
and $m^{2}=0$ (top-right), backreaction parameter $\kappa=0$ and
dimensionless quantity $\lambda=10$ in $d=5$ dimensions. As a matter
of fact, the other choices will not qualitatively modify our
results. From this figure, we can see the potential well forming in
all cases, which can trap the scalar particles. For the nonzero mass
of the scalar field, we observe that the potential well becomes
wider and deeper as the coupling parameter $\eta$ decreases, which
indicates that the increase of the coupling parameter will hinder
the condensate of the scalar field. For the case of $m^{2}=0$, we
find that the curves of the effective potential $V_{eff,0}(z)$
coincide, i.e., the Einstein tensor will not influence the effective
potential, which implies that the critical temperature is
independent of the Einstein tensor in this case. As we will show,
the behaviors of the effective potential are consistent with the
effects of the Einstein tensor on the condensate of the scalar
field. Considering that the effective scalar mass can give a better
shape in the potential as it was shown in \cite{KolyvarisKPSB}, we
also analyze the behaviors of the effective mass $m_{eff,0}(z)$ in
our holographic system
\begin{eqnarray}\label{meff0}
m_{eff,0}(z)=m^{2}-\left[1+\frac{(d-2)\eta
z^{2}f_{0}}{2r_{+}^{2}}\bigg(d-3-\frac{zf_{0}'}{f_{0}}\bigg)\right]\frac{\phi^{2}_{0}}{f_{0}},
\end{eqnarray}
which reduces to the standard effective mass given in Ref.
\cite{GubserPRD78} when $\eta\rightarrow0$. In Fig. \ref{VeffMeff},
we present the corresponding curves of the effective mass
$m_{eff,0}(z)$ with different values of the coupling parameter
$\eta$ for the fixed mass of the scalar field $m^{2}=-3$
(bottom-left) and $m^{2}=0$ (bottom-right), backreaction parameter
$\kappa=0$ and dimensionless quantity $\lambda=10$ in $d=5$
dimensions. Unfortunately, from Fig. \ref{VeffMeff}, we observe that
the behaviors of the effective mass are completely different from
those of the effective potential, i.e., the effective mass
(\ref{meff0}) can not give the correct behaviors of the effective
potential, which means that we have to count on the effective
potential in our models.

Near the asymptotic boundary $z=0$, we assume that $\psi_{1}$ takes
the form \cite{Siopsis}
\begin{eqnarray}\label{phiFz}
\psi_{1}(z)\sim \frac{\langle{\cal
O}_{i}\rangle}{r^{\Delta_{i}}_{+}} z^{\Delta_{i}}F(z),
\end{eqnarray}
where the trial function $F(z)$ obeys the boundary condition
$F(0)=1$. Substituting Eq. (\ref{phiFz}) into Eq. (\ref{psi1z}), we
arrive at
\begin{eqnarray}\label{Fzmotion}
\bigg[1+\frac{(d-2)\eta
z^{2}\xi}{2}\bigg(d-3-\frac{z\xi'}{\xi}\bigg)\bigg]F^{\prime\prime}+\bigg\{\frac{\xi'}{\xi}-\frac{d-4}{z}-
Q(z)&+&\nonumber \\ \frac{2\Delta_{i}}{z}\bigg[1+
\frac{(d-2)\eta z^{2}\xi}{2}\bigg(d-3-\frac{z\xi'}{\xi}\bigg)\bigg]\bigg\}F^{\prime}+\bigg\{\frac{\Delta_{i}(\Delta_{i}-1)}{z^{2}}\bigg[1+\frac{(d-2)\eta z^{2}\xi}{2}\bigg(d-3-\frac{z\xi'}{\xi}\bigg)\bigg]&+&\nonumber \\
\frac{\lambda^{2}(1-z^{d-3})^{2}}{z^{4}\xi^{2}}\bigg[1+\frac{(d-2)\eta
z^{2}\xi}{2}\bigg(d-3-\frac{z\xi'}{\xi}\bigg)\bigg]+\frac{\Delta_{i}}{z}\bigg[\frac{\xi'}{\xi}-\frac{d-4}{z}-Q(z)\bigg]-\frac{m^{2}}{z^{4}\xi}\bigg\}F&=&0,
\end{eqnarray}
with
\begin{eqnarray} Q(z)=\frac{(d-2)\eta z}{2}\left[(d-3)(d-6)
\xi-(3d-13)z\xi'+\frac{z^{2}\xi'^{2}}{\xi}+z^{2}\xi''\right].
\end{eqnarray}
In order to use the Sturm-Lioville method \cite{Siopsis}, we will
adopt an iteration method and express the backreaction parameter
$\kappa$ as $\kappa_{n}=n\Delta\kappa$ with
$n=0,1,2,\cdot\cdot\cdot$, where
$\Delta\kappa=\kappa_{n+1}-\kappa_{n}$ is the step size of our
iterative procedure. Setting $\kappa_{-1}=0$ and
$\lambda^{2}|_{\kappa_{-1}}=0$, we find that
$\kappa^{2}\lambda^{2}=\kappa_{n}^{2}\lambda^{2}=\kappa_{n}^{2}(\lambda^{2}|_{\kappa_{n-1}})+0[(\Delta\kappa)^{4}]$,
where $\lambda^{2}|_{\kappa_{n-1}}$ is the value of $\lambda^{2}$
for $\kappa_{n-1}$. Hence we can express the function $\xi(z)$ as
\begin{eqnarray}\label{xizNew}
\xi(z)\approx\frac{1}{z^{2}}-z^{d-3}-\frac{(d-3)\kappa_{n}^{2}(\lambda^{2}|_{\kappa_{n-1}})}{d-2}z^{d-3}(1-z^{d-3}).
\end{eqnarray}
Defining a new function
\begin{eqnarray}\label{TzNew}
T(z)=z^{4-d+2\Delta_{i}}\xi\bigg[1+\frac{(d-2)\eta
z^{2}\xi}{2}\bigg(d-3-\frac{z\xi'}{\xi}\bigg)\bigg],
\end{eqnarray}
we can rewrite Eq. (\ref{Fzmotion}) as
\begin{eqnarray}\label{TFzmotion}
(TF^{\prime})^{\prime}+z^{4-d+2\Delta_{i}}\xi\bigg\{\frac{\Delta_{i}(\Delta_{i}-1)}{z^{2}}\bigg[1+\frac{(d-2)\eta z^{2}\xi}{2}\bigg(d-3-\frac{z\xi'}{\xi}\bigg)\bigg]&+&\nonumber \\
\frac{\lambda^{2}(1-z^{d-3})^{2}}{z^{4}\xi^{2}}\bigg[1+\frac{(d-2)\eta
z^{2}\xi}{2}\bigg(d-3-\frac{z\xi'}{\xi}\bigg)\bigg]+\frac{\Delta_{i}}{z}\bigg(\frac{\xi'}{\xi}-\frac{d-4}{z}-Q\bigg)-\frac{m^{2}}{z^{4}\xi}\bigg\}F&=&0.
\end{eqnarray}
According to the Sturm-Liouville eigenvalue problem
\cite{Gelfand-Fomin}, we can deduce the eigenvalue $\lambda$
minimizes the expression
\begin{eqnarray}\label{SLEigenvalue}
\lambda^{2}=\frac{\int^{1}_{0}\left(TF'^{2}-UF^{2}\right)dz}{\int^{1}_{0}VF^{2}dz},
\end{eqnarray}
with
\begin{eqnarray}
U(z)&=&z^{4-d+2\Delta_{i}}\xi\bigg\{\frac{\Delta_{i}(\Delta_{i}-1)}{z^{2}}\bigg[1+\frac{(d-2)\eta z^{2}\xi}{2}\bigg(d-3-\frac{z\xi'}{\xi}\bigg)\bigg]+\frac{\Delta_{i}}{z}\bigg(\frac{\xi'}{\xi}-\frac{d-4}{z}-Q\bigg)-\frac{m^{2}}{z^{4}\xi}\bigg\},\nonumber \\
V(z)&=&\frac{z^{2\Delta_{i}-d}(1-z^{d-3})^{2}}{\xi}\bigg[1+\frac{(d-2)\eta
z^{2}\xi}{2}\bigg(d-3-\frac{z\xi'}{\xi}\bigg)\bigg].
\end{eqnarray}
Using Eq. (\ref{SLEigenvalue}) to calculate the minimum eigenvalue
of $\lambda^{2}$ for $i=+$ or $i=-$, we can obtain the critical
temperature $T_{c}$ for different coupling parameter $\eta$,
strength of the backreaction $\kappa$ and mass of the scalar field
$m$ from the following relation
\begin{eqnarray}\label{CTTc}
T_{c}=\frac{1}{4\pi}\left[(d-1)-\frac{(d-3)^{2}}{d-2}\kappa_{n}^{2}(\lambda^{2}|_{\kappa_{n-1}})\right]
\left(\frac{\rho}{\lambda}\right)^{\frac{1}{d-2}}.
\end{eqnarray}
For clarity, we focus on the condensate for the operator
$\langle{\cal O}_{+}\rangle$, just as mentioned in the previous
section. As a matter of fact, another choice for the operator
$\langle{\cal O}_{-}\rangle$ will not qualitatively modify our
results.

Before going further, we would like to make a comment. In order to
get the expression (\ref{SLEigenvalue}), we have used the following
boundary condition
\begin{eqnarray}\label{TFF}
[T(z)F(z)F'(z)]|_{0}^{1}=0.
\end{eqnarray}
Obviously, the condition $T(1)F(1)F'(1)=0$ can be satisfied easily
since we have $T(1)\equiv0$ from Eq. (\ref{TzNew}). On the other
hand, we observe that the leading order of $z$ in $T(z)$ near
$z\rightarrow0$ is $\beta=2-d+2\Delta_{+}$ in which $\beta\geq1$ for
$m^{2}\geq-(d-1)^{2}[2+(d-1)(d-2)\eta]/8$, which means that the
condition $T(0)F(0)F'(0)=0$ will be satisfied automatically. Thus,
we will just require $F(z)$ to satisfy the Dirichlet boundary
condition $F(0)=1$ rather than imposing the Neumann boundary
condition $F'(0)=0$, just as discussed in \cite{HFLi}. In the
following calculation, we will assume the trial function $F(z)$ to
be
\begin{eqnarray}
F(z)=1-az,
\end{eqnarray}
where $a$ is a constant. We find that it will give a better estimate
of the minimum of (\ref{SLEigenvalue}), which shows that the
analytical results are much more closer to the numerical findings.

As an example, we will calculate the case for $\eta=0$, $d=5$ and
$m^{2}=-3$ with the chosen values of the backreaction parameter
$\kappa$ for the operator $\langle{\cal O}_{+}\rangle$, i.e., $i=+$,
and compare with the analytical results in Ref. \cite{PanJWCh}.
Setting the step size $\Delta\kappa=0.05$, for $\kappa_{0}=0$ we
arrive at
\begin{eqnarray}
\lambda^{2}=\frac{120(\frac{3}{4}-\frac{9a}{7}-\frac{5a^{2}}{8})}{45+a(184-60\pi)
-60\log2 +20a^{2}(-2+\log8)},
\end{eqnarray}
whose minimum is $\lambda^{2}|_{\kappa_{0}}=17.5227$ at
$a=0.800597$. From Eq. (\ref{CTTc}), we can easily obtain the
critical temperature $T_{c}=0.197507\rho^{1/3}$, which is much
closer to the numerical result $T_{c}=0.197968\rho^{1/3}$
\cite{HorowitzPRD78}, compared with the analytical result
$T_{c}=0.196209\rho^{1/3}$ from the trial function $F(z)=1-az^{2}$
\cite{PanJWCh}. For $\kappa_{1}=0.05$, putting
$\lambda^{2}|_{\kappa_{0}}$ in Eqs. (\ref{xizNew}) and (\ref{TzNew})
we have
\begin{eqnarray}
\lambda^{2}=\frac{1.33528\times10^{2}(0.744524-1.27459a+0.619159a^{2})}{3.81409-5.03102a+1.77943a^{2}},
\end{eqnarray}
whose minimum is $\lambda^{2}|_{\kappa_{1}}=17.4250$ at
$a=0.798627$. So the critical temperature reads
$T_{c}=0.194805\rho^{1/3}$. Comparing with the analytical result
$T_{c}=0.193442\rho^{1/3}$ from Ref. \cite{PanJWCh}, we find that
this value is much closer to the numerical result
$T_{c}=0.195293\rho^{1/3}$. For $\kappa_{2}=0.10$, substituting
$\lambda^{2}|_{\kappa_{1}}$ into (\ref{xizNew}) and (\ref{TzNew}) we
get
\begin{eqnarray}
\lambda^{2}=\frac{2.28900\times10^{6}(0.728219-1.24146a+0.601767a^{2})}{6.63481\times10^{4}-8.77537a+3.11095\times10^{4}a^{2}},
\end{eqnarray}
whose minimum is $\lambda^{2}|_{\kappa_{2}}=17.1312$ at
$a=0.792406$. Therefore the critical temperature is
$T_{c}=0.186737\rho^{1/3}$, which is much closer to the numerical
finding $T_{c}=0.187414\rho^{1/3}$, compared with the analytical
result $T_{c}=0.185189\rho^{1/3}$ in \cite{PanJWCh}. For other
values of $\eta$, $\kappa$, $d$ and $m^{2}$, the similar iterative
procedure also can be applied to present the analytical result for
the critical temperature.

\begin{table}[ht]
\caption{\label{CriticalTcD5m3} The critical temperature $T_{c}$
obtained by the analytical method (left column) and numerical method
(right column) with the chosen values of the coupling parameter
$\eta$ and backreaction parameter $\kappa$ for the condensate of the
scalar operator in the case of 5-dimensional AdS black hole
background. Here we fix the mass of the scalar field by $m^{2}=-3$
and step size by $\Delta\kappa=0.05$.}
\begin{tabular}{c c c c}
         \hline\hline
$\kappa$ & 0  & 0.1 & 0.2
        \\
        \hline
~$\eta=-0.01$~~~~&~~~~$0.201355\rho^{1/3}$~~~$0.201740\rho^{1/3}$~~~~&~~~~$0.191578\rho^{1/3}$~~~$0.192137\rho^{1/3}$~~~~&~~~~$0.163579\rho^{1/3}$~~~$0.165238\rho^{1/3}$~
          \\
~$\eta=0$~~~~&~~~~$0.197507\rho^{1/3}$~~~$0.197968\rho^{1/3}$~~~~&~~~~$0.186737\rho^{1/3}$~~~$0.187414\rho^{1/3}$~~~~&~~~~$0.156051\rho^{1/3}$~~~$0.158049\rho^{1/3}$~
          \\
~$\eta=0.01$~~~~&~~~~$0.194565\rho^{1/3}$~~~$0.195092\rho^{1/3}$~~~~&~~~~$0.182953\rho^{1/3}$~~~$0.183736\rho^{1/3}$~~~~&~~~~$0.150009\rho^{1/3}$~~~$0.152312\rho^{1/3}$~
          \\
~$\eta=0.10$~~~~&~~~~$0.182643\rho^{1/3}$~~~$0.183528\rho^{1/3}$~~~~&~~~~$0.166684\rho^{1/3}$~~~$0.168092\rho^{1/3}$~~~~&~~~~$0.122372\rho^{1/3}$~~~$0.126425\rho^{1/3}$~
          \\
~$\eta=0.50$~~~~&~~~~$0.173569\rho^{1/3}$~~~$0.174842\rho^{1/3}$~~~~&~~~~$0.152918\rho^{1/3}$~~~$0.155093\rho^{1/3}$~~~~&~~~~$0.096657\rho^{1/3}$~~~$0.102711\rho^{1/3}$~
          \\
~$\eta=1.00$~~~~&~~~~$0.171502\rho^{1/3}$~~~$0.172879\rho^{1/3}$~~~~&~~~~$0.149553\rho^{1/3}$~~~$0.151950\rho^{1/3}$~~~~&~~~~$0.090000\rho^{1/3}$~~~$0.096586\rho^{1/3}$~
          \\
        \hline\hline
\end{tabular}
\end{table}

\begin{table}[ht]
\caption{\label{CriticalTcD4m2} The critical temperature $T_{c}$
obtained by the analytical method (left column) and numerical method
(right column) with the chosen values of the coupling parameter
$\eta$ and backreaction parameter $\kappa$ for the condensate of the
scalar operator in the case of 4-dimensional AdS black hole
background. Here we fix the mass of the scalar field by $m^{2}=-2$
and step size by $\Delta\kappa=0.05$.}
\begin{tabular}{c c c c}
         \hline\hline
$\kappa$ & 0  & 0.1 & 0.2
        \\
        \hline
~$\eta=-0.01$~~~~&~~~~$0.121567\rho^{1/2}$~~~$0.121763\rho^{1/2}$~~~~&~~~~$0.118937\rho^{1/2}$~~~$0.119154\rho^{1/2}$~~~~&~~~~$0.111255\rho^{1/2}$~~~$0.111646\rho^{1/2}$~
          \\
~$\eta=0$~~~~&~~~~$0.118197\rho^{1/2}$~~~$0.118426\rho^{1/2}$~~~~&~~~~$0.115330\rho^{1/2}$~~~$0.115582\rho^{1/2}$~~~~&~~~~$0.106975\rho^{1/2}$~~~$0.107435\rho^{1/2}$~
          \\
~$\eta=0.01$~~~~&~~~~$0.115561\rho^{1/2}$~~~$0.115822\rho^{1/2}$~~~~&~~~~$0.112487\rho^{1/2}$~~~$0.112777\rho^{1/2}$~~~~&~~~~$0.103553\rho^{1/2}$~~~$0.104076\rho^{1/2}$~
          \\
~$\eta=0.10$~~~~&~~~~$0.103856\rho^{1/2}$~~~$0.104318\rho^{1/2}$~~~~&~~~~$0.099591\rho^{1/2}$~~~$0.100111\rho^{1/2}$~~~~&~~~~$0.087345\rho^{1/2}$~~~$0.088302\rho^{1/2}$~
          \\
~$\eta=0.50$~~~~&~~~~$0.092740\rho^{1/2}$~~~$0.093527\rho^{1/2}$~~~~&~~~~$0.086689\rho^{1/2}$~~~$0.087606\rho^{1/2}$~~~~&~~~~$0.069614\rho^{1/2}$~~~$0.071322\rho^{1/2}$~
          \\
~$\eta=1.00$~~~~&~~~~$0.089693\rho^{1/2}$~~~$0.090597\rho^{1/2}$~~~~&~~~~$0.082976\rho^{1/2}$~~~$0.084046\rho^{1/2}$~~~~&~~~~$0.064117\rho^{1/2}$~~~$0.066112\rho^{1/2}$~
          \\
        \hline\hline
\end{tabular}
\end{table}

\begin{figure}[ht]
\includegraphics[scale=0.63]{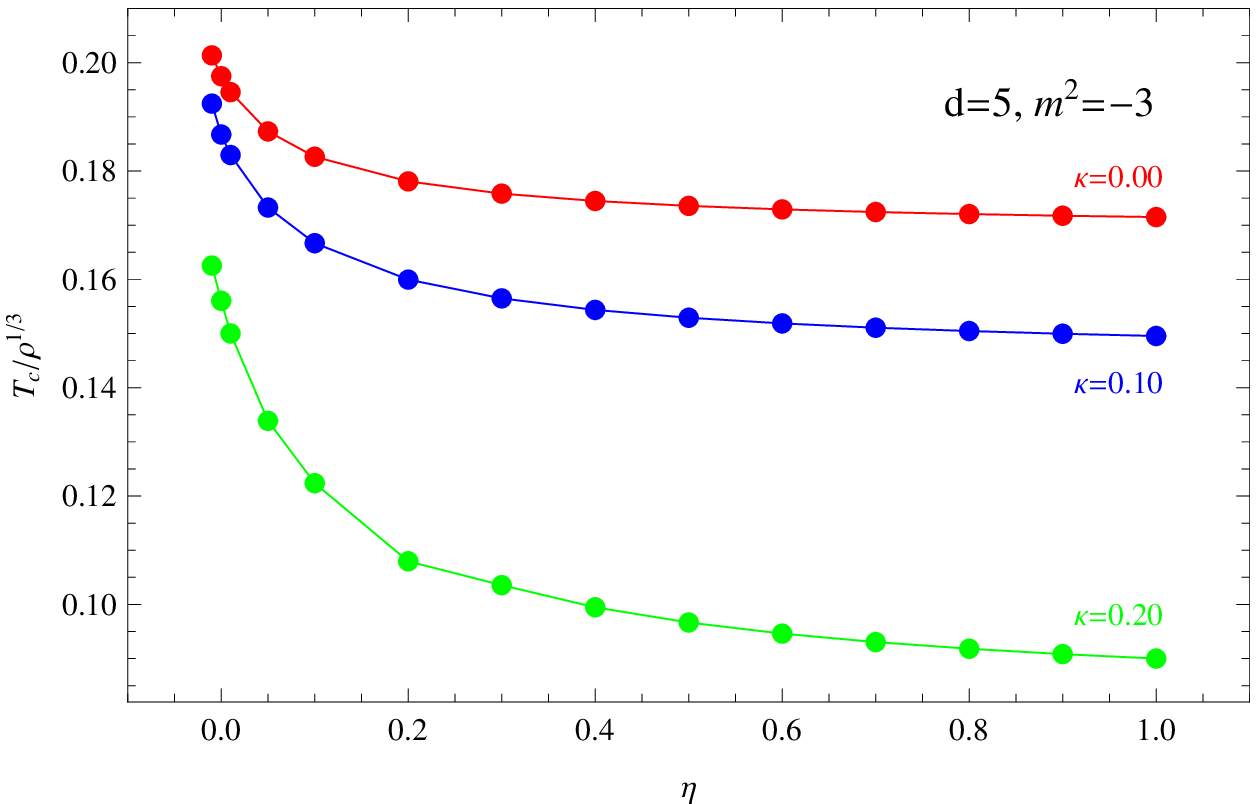}\hspace{0.2cm}%
\includegraphics[scale=0.63]{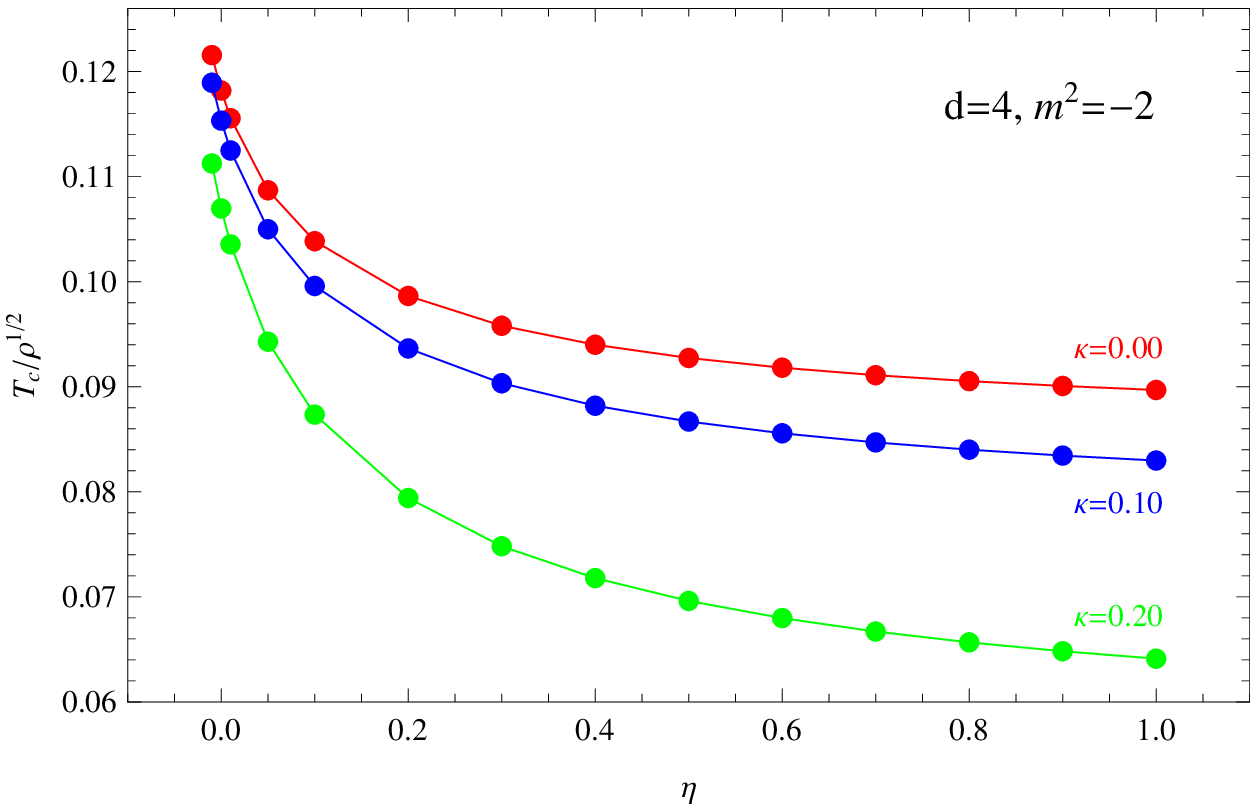}\\ \vspace{0.0cm}
\caption{\label{CriticalTemperatured5d4m32} (Color online) The
critical temperature $T_{c}$ obtained by the analytical method as a
function of the coupling parameter $\eta$ with the fixed mass of the
scalar field $m^{2}=-3$ in $d=5$ dimensions (left) and $m^{2}=-2$ in
$d=4$ dimensions (right). The three lines from top to bottom
correspond to increasing backreaction parameter $\kappa$, i.e.,
$\kappa=0.00$ (red), 0.10 (blue), and 0.20 (green), respectively.}
\end{figure}

In order to obtain the effect of the Einstein tensor on the critical
temperature for the scalar operator $\langle{\cal O}_{+}\rangle$, we
give the critical temperature $T_{c}$ obtained by the analytical
Sturm-Liouville method with the fixed masses of the scalar field by
$m^{2}=-3$ for the 5-dimensional AdS black hole and $m^{2}=-2$ for
the 4-dimensional one in Tables \ref{CriticalTcD5m3} and
\ref{CriticalTcD4m2}, respectively. In the calculation, we fix the
step size by $\Delta\kappa=0.05$. Moreover, to see the dependence of
the analytical results on the Einstein tensor more directly, in Fig.
\ref{CriticalTemperatured5d4m32} we also exhibit the critical
temperature $T_{c}$ obtained by the analytical method as a function
of the coupling parameter $\eta$ for the fixed backreaction
parameters and masses of the scalar field in $d=5$ (left) and $4$
(right) dimensions. For the fixed backreaction parameter, it is
clear that with the increase of the coupling parameter $\eta$, the
critical temperature $T_{c}$ decreases, which supports the
observation obtained in Ref. \cite{KuangE2016} and indicates that
the Einstein tensor will hinder the condensate of the scalar field.
Obviously, the effect of the Einstein tensor on the condensate of
the scalar field is consistent with the behavior of the effective
potential shown in Fig. \ref{VeffMeff}. On the other hand, imposing
the Dirichlet boundary condition of the trial function $F(z)$
without the Neumann boundary conditions, we observe that the
improved Sturm-Liouville method can indeed give a better estimate of
the critical temperature, compared with the analytical result from
the trial function $F(z)=1-az^{2}$ in Ref. \cite{PanJWCh}.

\begin{table}[ht]
\caption{\label{CriticalTcD5m0} The critical temperature $T_{c}$
obtained by the analytical method (left column) and numerical method
(right column) with the chosen values of the coupling parameter
$\eta$ and backreaction parameter $\kappa$ for the condensate of the
scalar operator in the case of 5-dimensional AdS black hole
background. Here we fix the mass of the scalar field by $m^{2}=0$
and step size by $\Delta\kappa=0.05$.}
\begin{tabular}{c c c c}
         \hline\hline
$\kappa$ & 0  & 0.1 & 0.2
        \\
        \hline
~$\eta=-0.01$~~~~&~~~~$0.168998\rho^{1/3}$~~~$0.170509\rho^{1/3}$~~~~&~~~~$0.145743\rho^{1/3}$~~~$0.148561\rho^{1/3}$~~~~&~~~~$0.085189\rho^{1/3}$~~~$0.093255\rho^{1/3}$~
          \\
~$\eta=0$~~~~&~~~~$0.168998\rho^{1/3}$~~~$0.170509\rho^{1/3}$~~~~&~~~~$0.145719\rho^{1/3}$~~~$0.148530\rho^{1/3}$~~~~&~~~~$0.084974\rho^{1/3}$~~~$0.093006\rho^{1/3}$~
          \\
~$\eta=0.01$~~~~&~~~~$0.168998\rho^{1/3}$~~~$0.170509\rho^{1/3}$~~~~&~~~~$0.145697\rho^{1/3}$~~~$0.148502\rho^{1/3}$~~~~&~~~~$0.084782\rho^{1/3}$~~~$0.092783\rho^{1/3}$~
          \\
~$\eta=0.10$~~~~&~~~~$0.168998\rho^{1/3}$~~~$0.170509\rho^{1/3}$~~~~&~~~~$0.145576\rho^{1/3}$~~~$0.148345\rho^{1/3}$~~~~&~~~~$0.083683\rho^{1/3}$~~~$0.091482\rho^{1/3}$~
          \\
~$\eta=0.50$~~~~&~~~~$0.168998\rho^{1/3}$~~~$0.170509\rho^{1/3}$~~~~&~~~~$0.145432\rho^{1/3}$~~~$0.148158\rho^{1/3}$~~~~&~~~~$0.082343\rho^{1/3}$~~~$0.089839\rho^{1/3}$~
          \\
~$\eta=1.00$~~~~&~~~~$0.168998\rho^{1/3}$~~~$0.170509\rho^{1/3}$~~~~&~~~~$0.145391\rho^{1/3}$~~~$0.148104\rho^{1/3}$~~~~&~~~~$0.081952\rho^{1/3}$~~~$0.089346\rho^{1/3}$~
          \\
        \hline\hline
\end{tabular}
\end{table}

\begin{table}[ht]
\caption{\label{CriticalTcD4m0} The critical temperature $T_{c}$
obtained by the analytical method (left column) and numerical method
(right column) with the chosen values of the coupling parameter
$\eta$ and backreaction parameter $\kappa$ for the condensate of the
scalar operator in the case of 4-dimensional AdS black hole
background. Here we fix the mass of the scalar field by $m^{2}=0$
and step size by $\Delta\kappa=0.05$.}
\begin{tabular}{c c c c}
         \hline\hline
$\kappa$ & 0  & 0.1 & 0.2
        \\
        \hline
~$\eta=-0.01$~~~~&~~~~$0.085581\rho^{1/2}$~~~$0.086667\rho^{1/2}$~~~~&~~~~$0.078022\rho^{1/2}$~~~$0.079368\rho^{1/2}$~~~~&~~~~$0.057473\rho^{1/2}$~~~$0.060125\rho^{1/2}$~
          \\
~$\eta=0$~~~~&~~~~$0.085581\rho^{1/2}$~~~$0.086667\rho^{1/2}$~~~~&~~~~$0.078015\rho^{1/2}$~~~$0.079360\rho^{1/2}$~~~~&~~~~$0.057433\rho^{1/2}$~~~$0.060080\rho^{1/2}$~
          \\
~$\eta=0.01$~~~~&~~~~$0.085581\rho^{1/2}$~~~$0.086667\rho^{1/2}$~~~~&~~~~$0.078009\rho^{1/2}$~~~$0.079353\rho^{1/2}$~~~~&~~~~$0.057394\rho^{1/2}$~~~$0.060037\rho^{1/2}$~
          \\
~$\eta=0.10$~~~~&~~~~$0.085581\rho^{1/2}$~~~$0.086667\rho^{1/2}$~~~~&~~~~$0.077963\rho^{1/2}$~~~$0.079303\rho^{1/2}$~~~~&~~~~$0.057125\rho^{1/2}$~~~$0.059734\rho^{1/2}$~
          \\
~$\eta=0.50$~~~~&~~~~$0.085581\rho^{1/2}$~~~$0.086667\rho^{1/2}$~~~~&~~~~$0.077886\rho^{1/2}$~~~$0.079211\rho^{1/2}$~~~~&~~~~$0.056619\rho^{1/2}$~~~$0.059156\rho^{1/2}$~
          \\
~$\eta=1.00$~~~~&~~~~$0.085581\rho^{1/2}$~~~$0.086667\rho^{1/2}$~~~~&~~~~$0.077839\rho^{1/2}$~~~$0.079173\rho^{1/2}$~~~~&~~~~$0.056408\rho^{1/2}$~~~$0.058911\rho^{1/2}$~
          \\
        \hline\hline
\end{tabular}
\end{table}

\begin{figure}[ht]
\includegraphics[scale=0.466]{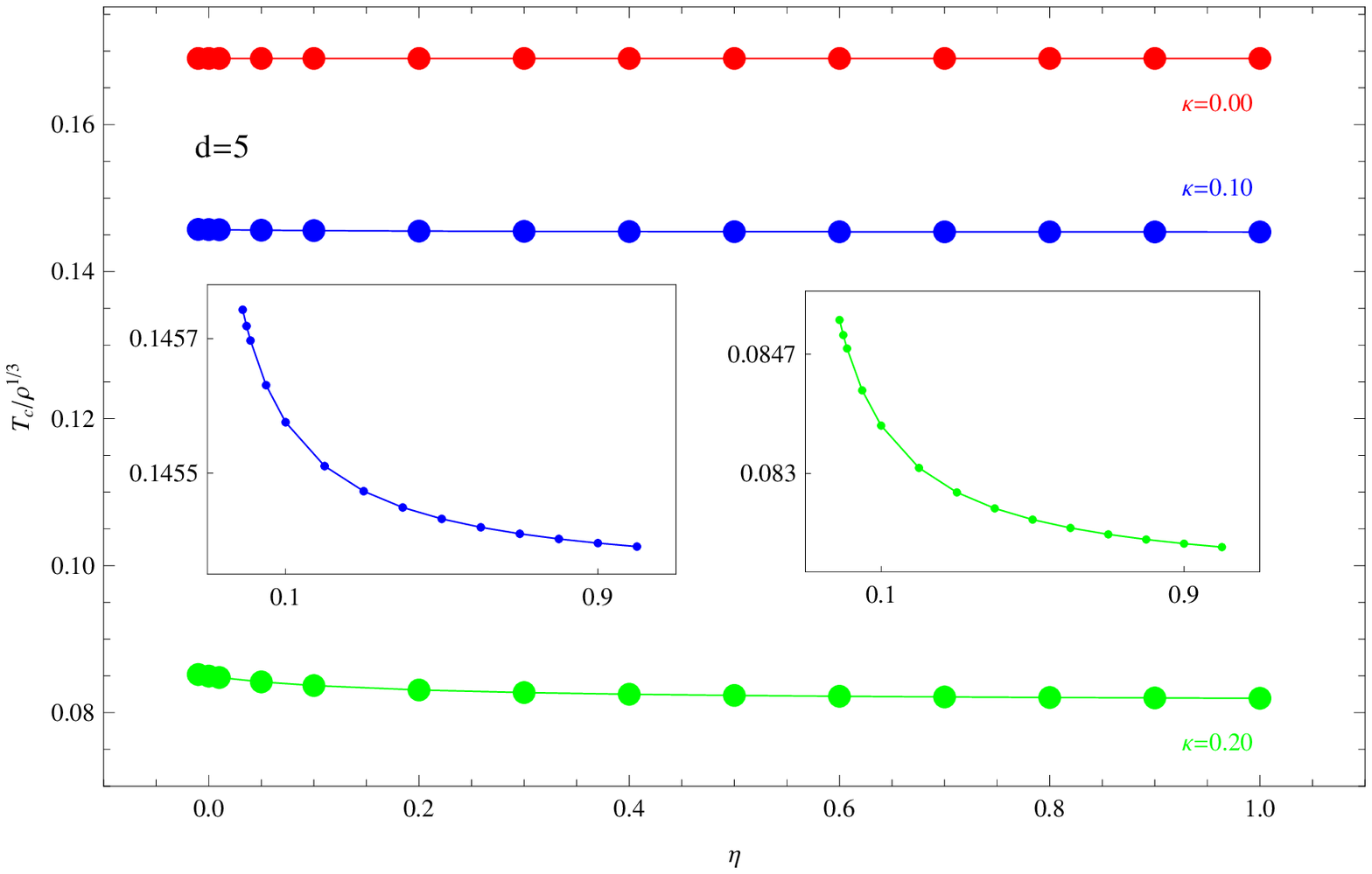}\hspace{0.2cm}%
\includegraphics[scale=0.4]{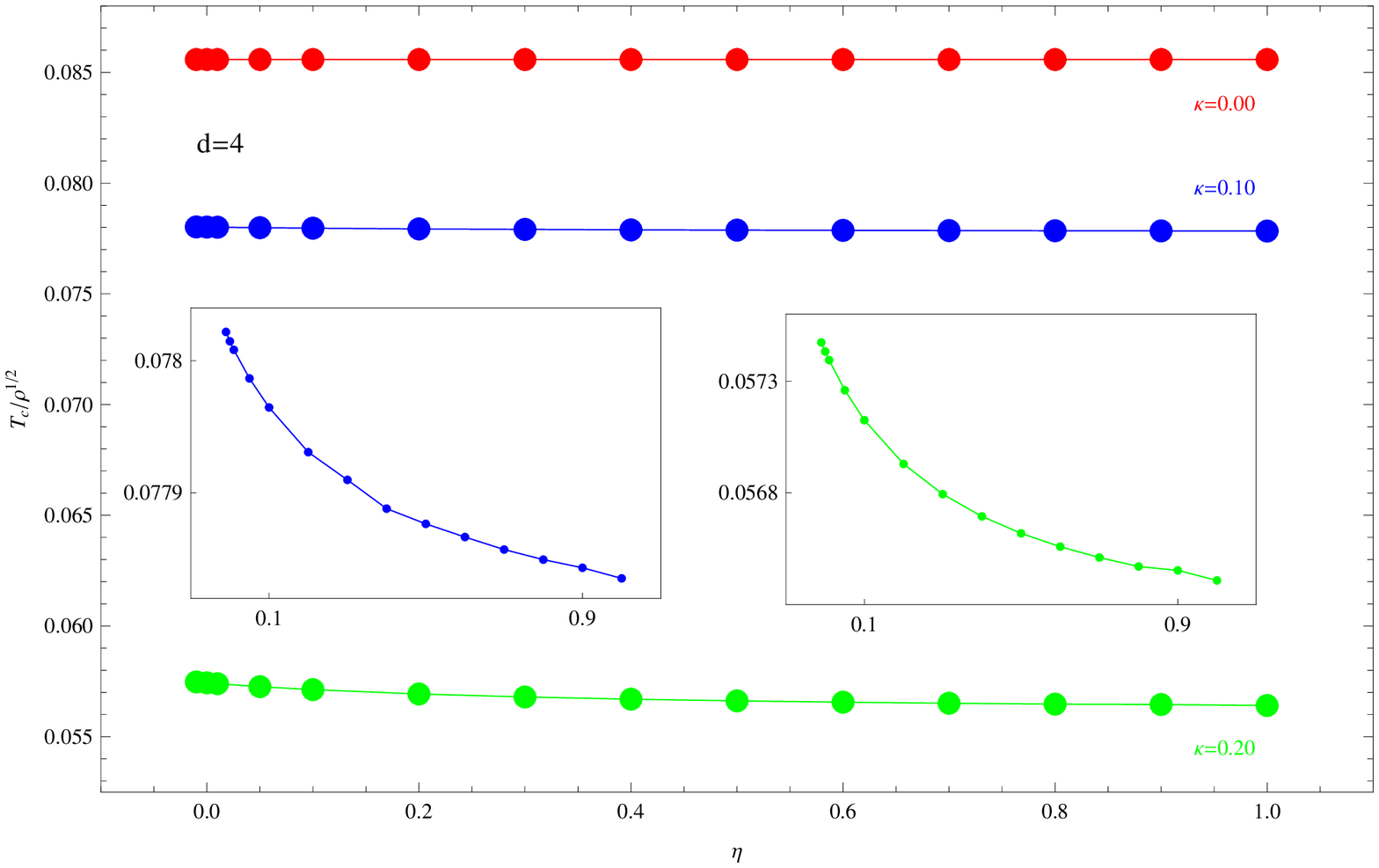}\\ \vspace{0.0cm}
\caption{\label{CriticalTemperatured5d4m0} (Color online) The
critical temperature $T_{c}$ obtained by the analytical method as a
function of the coupling parameter $\eta$ with the fixed mass of the
scalar field $m^{2}=0$ in $d=5$ (left) and $4$ (right) dimensions.
The three lines from top to bottom correspond to increasing
backreaction parameter $\kappa$, i.e., $\kappa=0.00$ (red), 0.10
(blue), and 0.20 (green), respectively.}
\end{figure}

Considering the fact that the effect of the Einstein tensor is
intertwined with that of the mass of the scalar field in the
expression (\ref{CharacteristicExponent}), we will set $m^{2}=0$ to
get the pure effect of the Einstein tensor on the critical
temperature. In Tables \ref{CriticalTcD5m0}, \ref{CriticalTcD4m0}
and Fig. \ref{CriticalTemperatured5d4m0}, we provide the critical
temperature $T_{c}$ obtained by the analytical Sturm-Liouville
method with the fixed mass of the scalar field by $m^{2}=0$ and step
size by $\Delta\kappa=0.05$ for the 5- and 4-dimensional AdS black
hole backgrounds respectively. For the fixed nonzero parameter
$\kappa$, the conclusion still holds, i.e., the Einstein tensor
hinders the condensate of the scalar field. However, for the probe
limit $\kappa=0$, from the leftmost columns in Tables
\ref{CriticalTcD5m0}, \ref{CriticalTcD4m0} and the red lines in Fig.
\ref{CriticalTemperatured5d4m0} we find that $T_{c}$ is independent
of the coupling parameter $\eta$, which implies that the Einstein
tensor will not affect the condensate of the scalar field for the
case of $m^{2}=0$. Again, in this case the effect of the Einstein
tensor on the condensate of the scalar field agrees well with the
behavior of the effective potential shown in Fig. \ref{VeffMeff}.
Thus, the probe approximation loses some important information and
we have to count on the backreaction to explore the real impact of
the Einstein tensor on the holographic superconductors in this case.
Moreover, in the case of $m^{2}=0$, we observe that the critical
temperature $T_{c}$ increases as the spacetime dimension $d$
increases for the fixed coupling parameter $\eta$ and backreaction
parameter $\kappa$, which supports the findings in Ref.
\cite{Pan-WangGB2010} and means that the increase of the
dimensionality of the AdS space makes it easier for the scalar hair
to be formed.

On the other hand, from Tables
\ref{CriticalTcD5m3}-\ref{CriticalTcD4m0} and Figs.
\ref{CriticalTemperatured5d4m32}-\ref{CriticalTemperatured5d4m0}, we
point out that the critical temperature $T_{c}$ decreases as the
backreaction parameter $\kappa$ increases for the fixed coupling
parameter $\eta$, scalar field mass $m^{2}$ and spacetime dimension
$d$, which shows that the stronger backreaction can make the scalar
hair more difficult to be developed. This can be used to back up the
findings in Refs.
\cite{PanJWCh,BarclayGregory,Barclay2011,GregoryRev,KannoCQG,CSJHEP2015}.

\section{Critical behavior from marginally stable modes}

We now use the shooting method
\cite{HartnollRev,HerzogRev,HorowitzRev,CaiRev} to study the
marginally stable modes of the scalar perturbation coupled to the
electric field and Einstein tensor, which can reveal the critical
behavior of the holographic superconductors with backreactions from
the coupling of a scalar field to the Einstein tensor near the phase
transition point numerically. We will also compare this numerical
result with the analytical one in order to test the effectiveness
and accuracy of the Sturm-Liouville method.

Considering the scalar perturbation in the Reissner-Nordstr\"{o}m
AdS black hole background (\ref{RNBH}) coupled to a Maxwell field
and Einstein tensor, from the action (\ref{System}) we can get the
Euler-Lagrange equation of motion for the perturbation field
\begin{eqnarray}\label{eompsi}
\frac{1}{\sqrt{-g}}(\partial_\mu-iA_\mu)\left[\sqrt{-g}(g^{\mu\nu}+\eta
G^{\mu\nu})(\partial_\nu-iA_\nu)\right]\tilde{\psi}-m^2\tilde{\psi}=0,
\end{eqnarray}
with the nonzero components of the Einstein tensor
\begin{eqnarray}\label{RNEinsteinTensor}
&&G^{tt}=-\frac{d-2}{2r^{2}}\left[(d-3)+\frac{rf^{\prime}}{f}\right],~~
G^{rr}=\frac{(d-2)f^{2}}{2r^{2}}\left[(d-3)+\frac{rf^{\prime}}{f}\right],\nonumber \\
&&G^{xx}=G^{yy}=\cdot\cdot\cdot=\frac{(d-3)(d-4)f+r[2(d-3)f^{\prime}+rf^{\prime\prime}]}{2r^{4}},
\end{eqnarray}
where $f$ is given by (\ref{RNBH}). Assuming
$\tilde{\psi}=e^{-i\omega t}R(r)Y(x^{i})$ and making the separation
of the variables, we obtain
\begin{eqnarray}\label{Reom}
\left\{1+\frac{(d-2)\eta}{2r}\bigg[f'+\frac{(d-3)f}{r}\bigg]\right\}R^{\prime\prime}+\bigg\{\left(\frac{d-2}{r}+\frac{f'}{f}\right)+\frac{(d-2)\eta}{2r}&\times&\nonumber\\
\bigg[f''+\frac{3(d-3)f'}{r}+\frac{f'^{2}}{f}+\frac{(d-3)(d-4)f}{r^{2}}\bigg]\bigg\}R^\prime+
\frac{(\phi+\omega)^{2}}{f^{2}}\left\{1+\frac{(d-2)\eta}{2r}\bigg[f'+\frac{(d-3)f}{r}\bigg]\right\}R&-&\nonumber\\
\frac{A_{n}^{2}}{r^{2}f}\left\{1+\frac{\eta}{2}\bigg[f''+\frac{2(d-3)f'}{r}+\frac{(d-3)(d-4)f}{r^{2}}\bigg]\right\}R-\frac{m^{2}}{f}R&=&0,
\end{eqnarray}
where $\phi$ has been introduced in (\ref{RNBH}) and $A_{n} \in
\mathbb{Z}$ is the eigenvalue of the following equation
\begin{eqnarray}
\Sigma_{i}\left[\frac{\partial^{2}Y(x^{i})}{\partial
x^{i2}}\right]+A_{n}^{2}\Sigma_{i}Y(x^{i})=0,
\end{eqnarray}
with $x^i=x,y,\cdots$. It is expected that the lowest mode $A_{n}=0$
will be the first to condense and result in the most stable solution
after condensing, which means that there are no momenta in the
$(x,y,...)$-directions \cite{marginally stable modes}. Then,
changing the variable from $r$ to $z=r_{+}/r$ for convenience, we
have the equation of motion
\begin{eqnarray}\label{Rzeom}
\left\{1+\frac{(d-2)\eta
z^{2}f}{2r_{+}^{2}}\bigg[(d-3)-\frac{zf'}{f}\bigg]\right\}R^{\prime\prime}
-\bigg\{\left(\frac{d-4}{z}-\frac{f'}{f}\right)+\frac{(d-2)\eta
z}{2r_{+}^{2}}\bigg[z^{2}f''-(3d-13)zf'&+&\nonumber\\
\frac{z^{2}f'^{2}}{f}+(d-3)(d-6)f\bigg]\bigg\}R^\prime+
\frac{r_{+}^{2}(\phi+\omega)^{2}}{z^{4}f^{2}}\bigg\{1+\frac{(d-2)\eta
z^{2}f}{2r_{+}^{2}}\bigg[(d-3)-\frac{zf'}{f}\bigg]\bigg\}R-\frac{m^{2}r_{+}^{2}}{z^{4}f}R&=&0,
\end{eqnarray}
where the prime now denotes the derivative with respect to $z$.

It is well known that the marginally stable modes correspond to
$\omega=0$ which indicates that the phase transition or the critical
phenomena may occur \cite{GubserPRD78}. Thus, we will solve the
equation of motion \eqref{Rzeom} numerically by doing integration
from the horizon out to the infinity in the case of $\omega=0$ with
the boundary conditions of $R(z)$ at the event horizon
\begin{eqnarray}\label{rzzEvent}
R(z)=R(1)-R'(1)(1-z)+\frac{1}{2}R''(1)(1-z)^{2}+\cdots\,,
\end{eqnarray}
and at the asymptotic AdS boundary
\begin{eqnarray}\label{rzzBoundary}
R(z)=\frac{R_{-}}{r^{\Delta_{-}}_{+}}z^{\Delta_{-}}+\frac{R_{+}}{r_{+}^{\Delta_{+}}}z^{\Delta_{+}}.
\end{eqnarray}
Since we concentrate on the condensate for the operator
$\langle{\cal O}_{+}\rangle$ in this work, we impose boundary
condition $R_{-}=0$. In the following calculations, we will scan the
parameter space of holographic superconductors and find the certain
values of $\lambda=\rho/r^{d-2}_{+c}$ which satisfy the boundary
condition for the given $\eta$, $\kappa$, $m^{2}$ and $d$. Note that
the quantity of $R(1)$ is very close to zero near the critical point
of the phase transition, we set the initial condition $R(1)=0.001$
without loss of generality.

\begin{figure}[ht]
\includegraphics[scale=0.42]{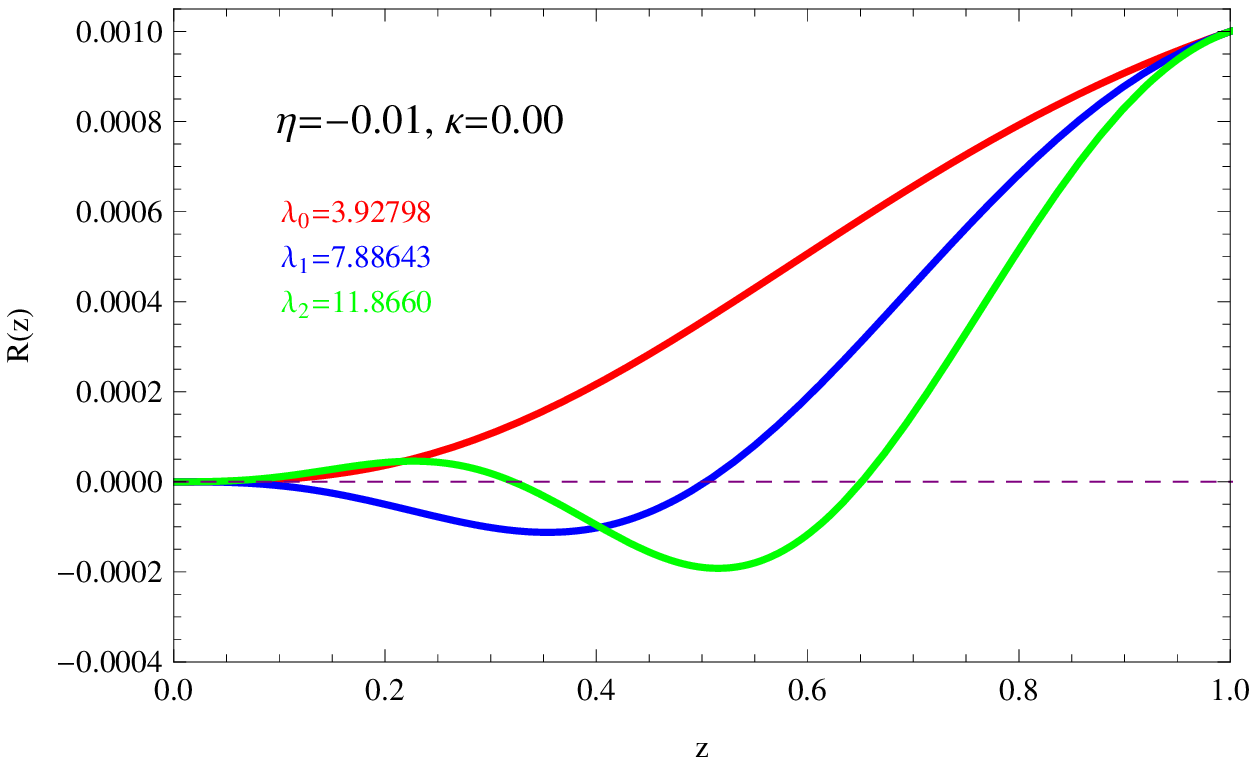}\hspace{0.2cm}%
\includegraphics[scale=0.42]{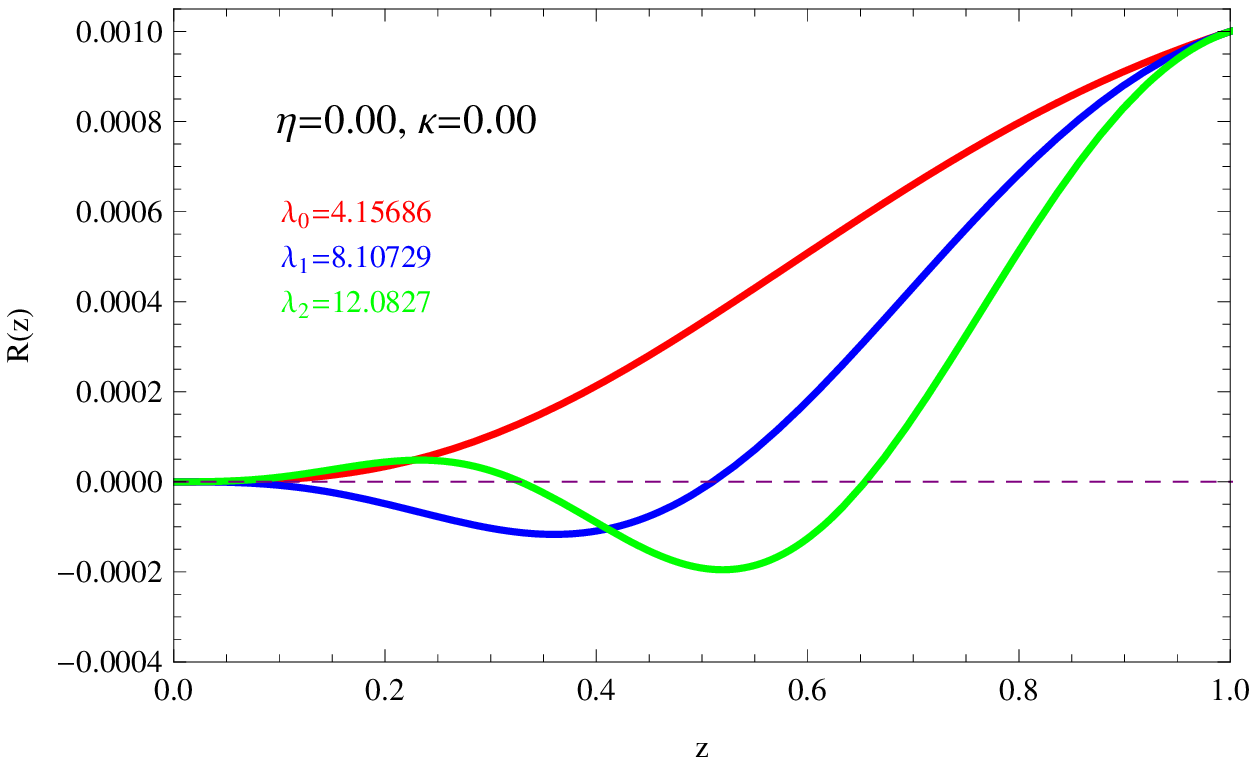}\hspace{0.2cm}%
\includegraphics[scale=0.42]{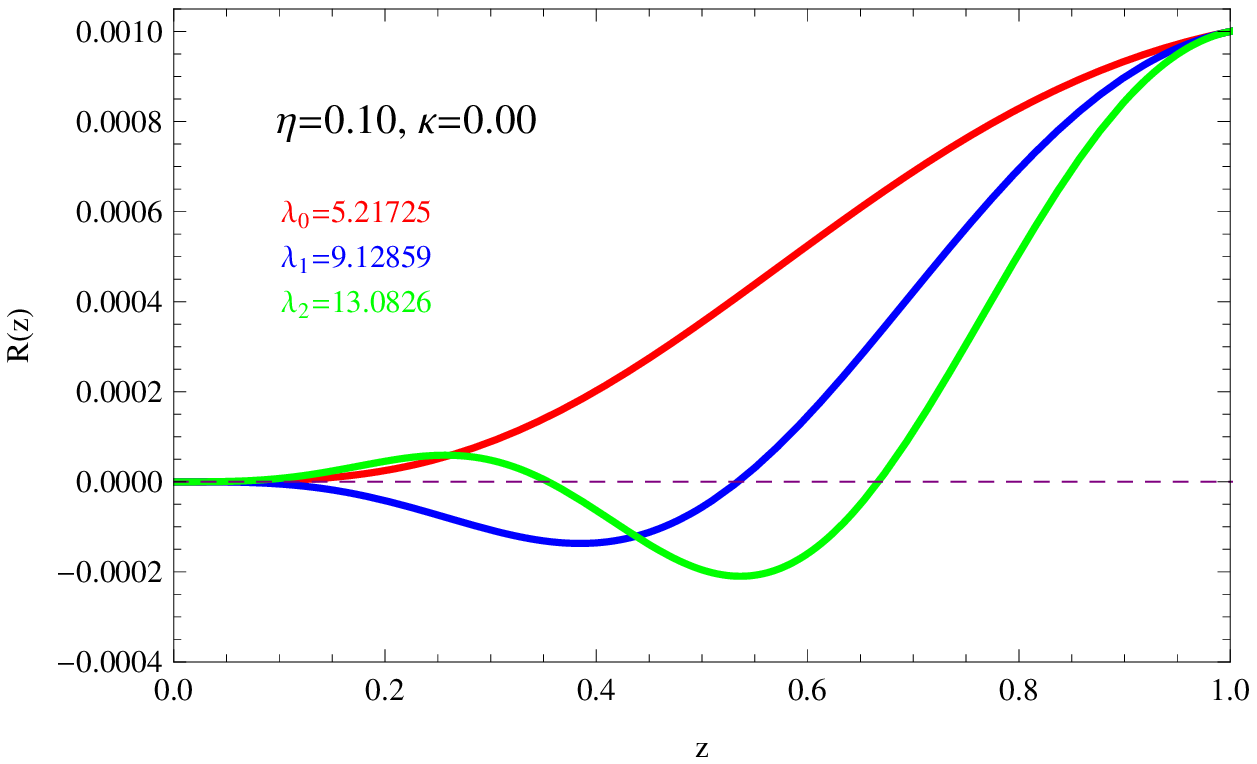}\\ \vspace{0.0cm}
\includegraphics[scale=0.42]{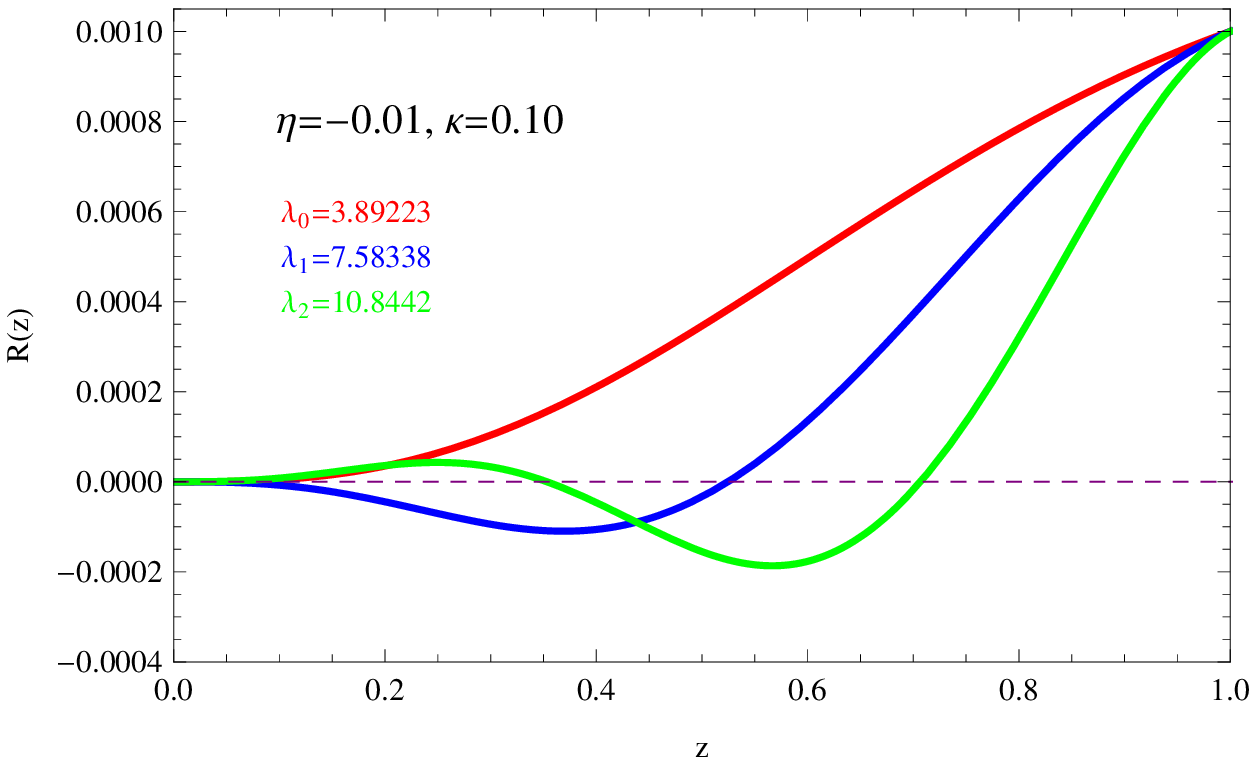}\hspace{0.2cm}%
\includegraphics[scale=0.42]{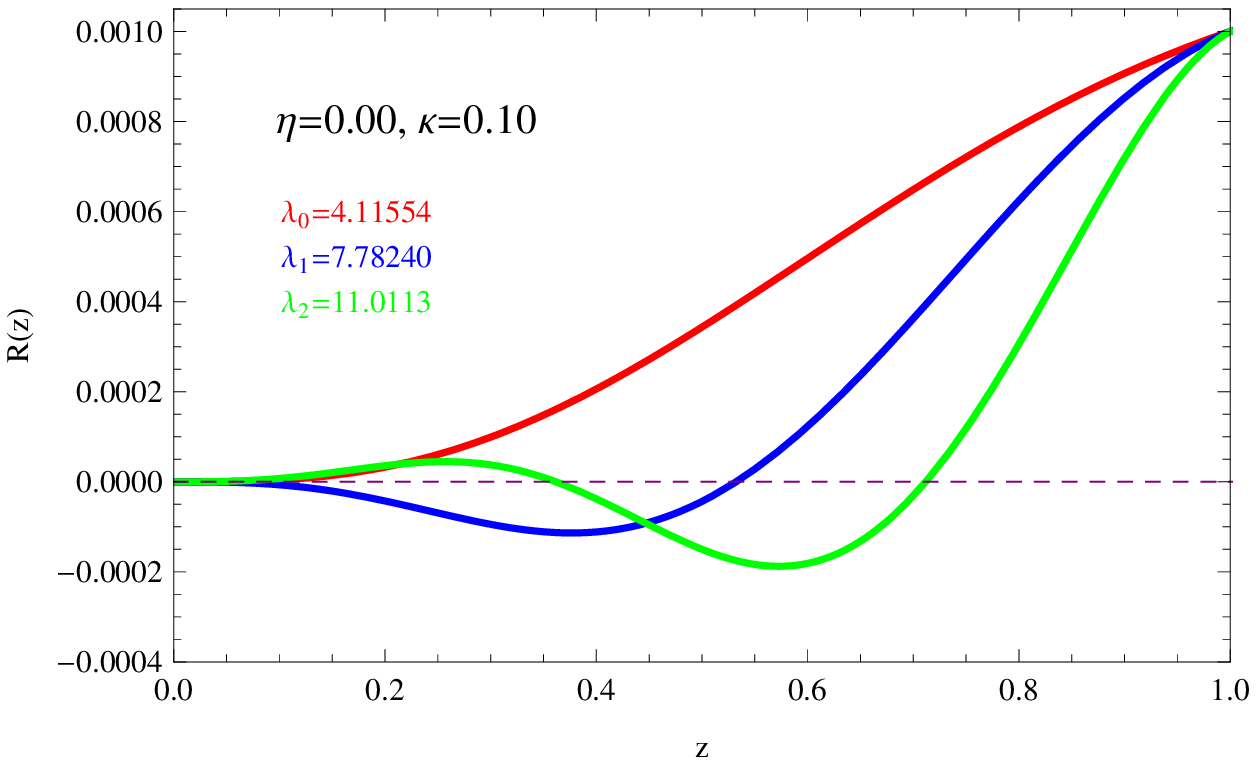}\hspace{0.2cm}%
\includegraphics[scale=0.42]{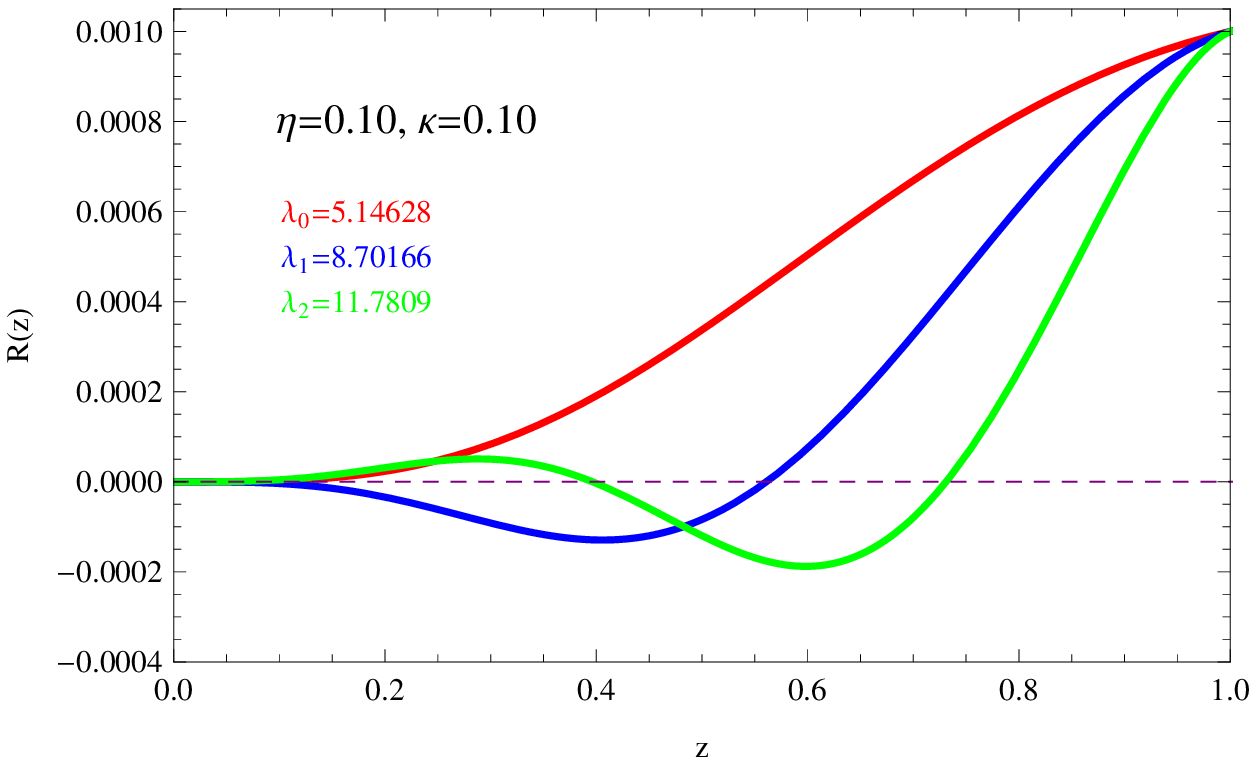}\\ \vspace{0.0cm}
\includegraphics[scale=0.42]{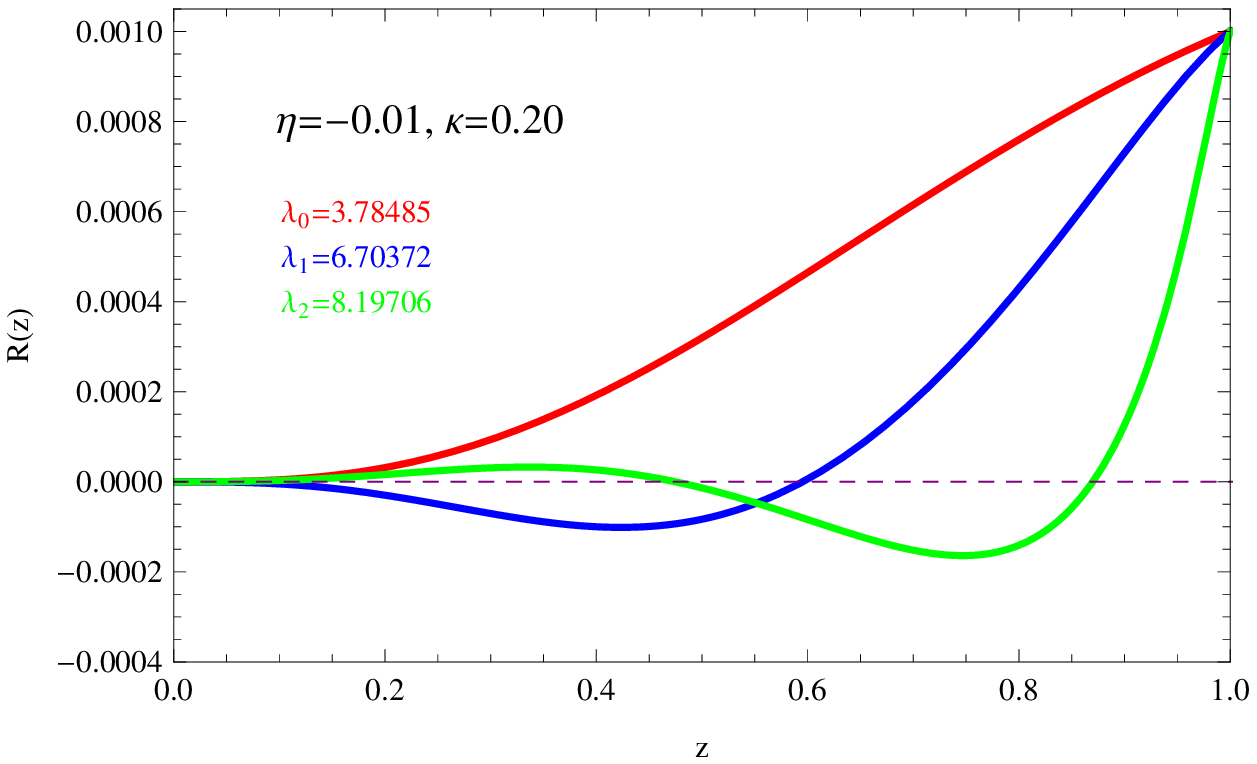}\hspace{0.2cm}%
\includegraphics[scale=0.42]{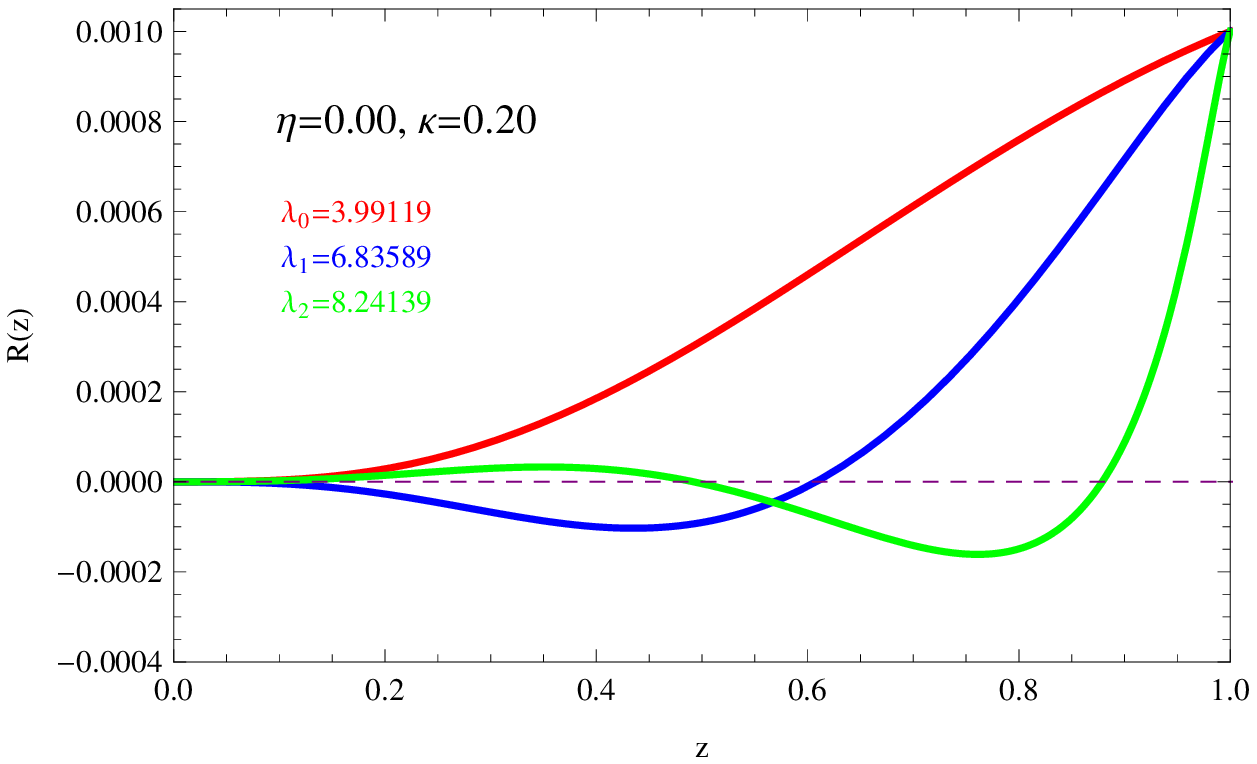}\hspace{0.2cm}%
\includegraphics[scale=0.42]{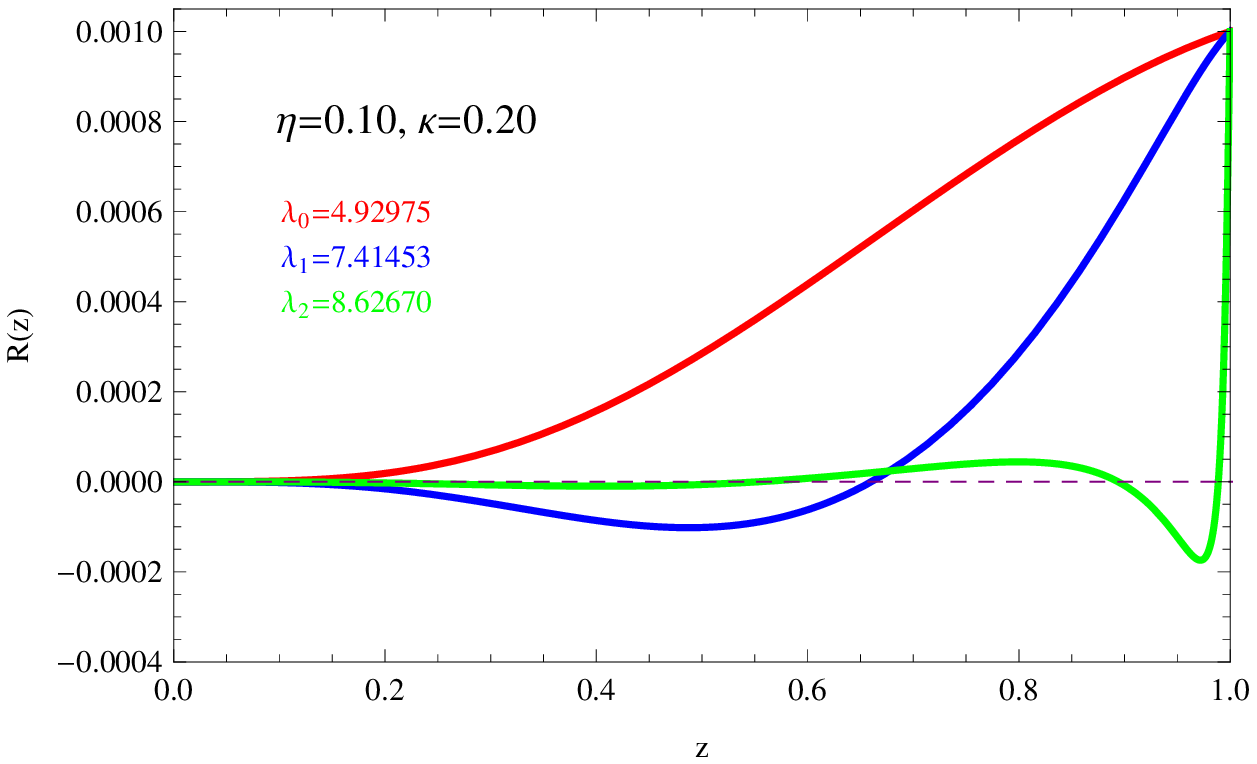}\\ \vspace{0.0cm}
\caption{\label{notes} (Color online.) The marginally stable curves
of scalar fields corresponding to various critical values
$\lambda_{n}$ with $m^{2}=-3$ for different coupling parameters
$\eta$ and backreaction parameters $\kappa$ in the case of
5-dimensional AdS black hole background. The first three
lowest-lying critical values $\lambda_{n}$ for different curves are
$\lambda_{0}<\lambda_{1}<\lambda_{2}$ in the sequence.}
\end{figure}

In Fig. \ref{notes}, we plot the marginally stable curves of scalar
fields $R(z)$ corresponding to the critical values $\lambda_{n}$
with $m^{2}=-3$ for different coupling parameters $\eta$ and
backreaction parameters $\kappa$ in the case of 5-dimensional AdS
black hole background by solving Eq. (\ref{Rzeom}) numerically. In
each panel, three curves correspond to the first three lowest-lying
critical values $\lambda_{n}$ which are
$\lambda_{0}<\lambda_{1}<\lambda_{2}$ in the sequence, where the
index $n$ denotes the ``overtone number" \cite{marginally stable
modes}. The red line has no intersecting points with the $R(z)=0$
axis at nonvanishing $z$ and is dual to the minimal value of
$\lambda_{n}$ (a mode of node $n=0$), which will be the first to
condense. The blue line (a mode of node $n=1$) has one intersecting
point with $R(z)=0$ axis while the green line (a mode of node $n=2$)
has two, which do not matter to the phase transition because the
blue and green lines are expected to be unstable with radial
oscillations in $z$-direction of $R(z)$ costing energy
\cite{GubserPufu}. At the critical point, inserting $\lambda_{0}$
into the Hawking temperature of the $d$-dimensional
Reissner-Nordstr\"{o}m black hole (\ref{RNBH}), i.e.,
\begin{eqnarray}\label{RNHawkingTemperature}
T_{H}=\frac{r_{+}}{4\pi}\left[(d-1)-\frac{(d-3)^{2}\kappa^{2}}{d-2}\left(\frac{\rho}{r^{d-2}_{+}}\right)^{2}\right],
\end{eqnarray}
we can easily obtain the critical temperature $T_c$. For example,
for the case of $\eta=0.00,~\kappa=0.10$ with $m^{2}=-3$ and $d=5$,
from Fig. \ref{notes} we have $\lambda_{0}=4.11554$, which leads to
$T_{c}=0.187414\rho^{1/3}$. Similarly, we can get the critical
temperatures for different values of $\eta$, $\kappa$, $d$ and
$m^{2}$.

In Tables \ref{CriticalTcD5m3}-\ref{CriticalTcD4m0}, we present the
critical temperatures obtained numerically by using the shooting
method for the 5-dimensional and 4-dimensional black hole
backgrounds, respectively. Compared with the analytical results in
each table, the agreement of the numerical calculation (right
column) and analytical result derived from the Sturm-Liouville
method (left column) is impressive, which implies that the
Sturm-Liouville method is still powerful to study the holographic
superconductors from the coupling of a scalar field to the Einstein
tensor even if we consider the backreactions. Obviously, the
``marginally stable modes" method is a very effective way to study
the critical behavior of the phase transition in the holographic
superconductor models. In addition to giving us the numerical
results of the critical temperature, the marginally stable modes can
reveal the instabilities of the background which means that the AdS
black hole will become unstable to develop charged scalar hairs in
the AdS black hole background.

\section{Conclusions}

We have investigated the properties of the backreacting holographic
superconductors from the coupling of a scalar field to the Einstein
tensor in the background of a $d$-dimensional AdS black hole, which
provides a more explicit and complete understanding of the effect of
the Einstein tensor on the holographic superconductors. Imposing the
Dirichlet boundary condition of the trial function $F(z)$ without
the Neumann boundary conditions, we improved the analytical
Sturm-Liouville method with an iterative procedure to calculate the
critical temperatures for the scalar operator $\langle{\cal
O}_{+}\rangle$ and found that the analytical findings obtained in
this way are in very good agreement with the numerical results from
the ``marginally stable modes" method, which implies that the
Sturm-Liouville method is still powerful to study the holographic
superconductors from the coupling of a scalar field to the Einstein
tensor even if we consider the backreactions. It is shown that, when
the backreaction parameter is nonzero, the critical temperature
decreases with the increase of the coupling parameter of the
Einstein tensor, which can be used to back up the observation
obtained in Ref. \cite{KuangE2016} that the Einstein tensor will
hinder the condensate of the scalar field. However, when the
backreaction parameter and scalar mass are zero, the critical
temperature is independent of the Einstein tensor, which implies
that the probe approximation still loses some important information
and we have to count on the backreaction to explore the real and
pure impact of the Einstein tensor on the holographic
superconductors in this case. In addition, we observed that the
critical temperature increases as the spacetime dimension increases
for the fixed scalar mass, coupling parameter and backreaction
parameter, which means that the scalar hair can be formed easier in
the higher-dimensional background. Moreover, we interestingly noted
that the Einstein tensor, backreaction and spacetime dimension
cannot modify the critical phenomena, i.e., this holographic
superconductor phase transition belongs to the second order and the
critical exponent of the system always takes the mean-field value.

\begin{acknowledgments}

We thank Professor Eleftherios Papantonopoulos for his helpful
discussions and suggestions. This work was supported by the National
Natural Science Foundation of China under Grant Nos. 11775076,
11475061 and 11690034; Hunan Provincial Natural Science Foundation
of China under Grant No. 2016JJ1012.

\end{acknowledgments}


\begin{thebibliography}{99}

\bibitem{Tinkham}
M. Tinkham, Introduction to Superconductivity, 2nd ed.(McGrawHill,
New York, 1996).

\bibitem{BCS}
J. Bardeen, L.N. Cooper, and J.R. Schrieffer, Phys. Rev. {\bf 108},
1175 (1957).

\bibitem{Maldacena}
J. Maldacena,
Adv. Theor. Math. Phys. {\bf 2}, 231 (1998) [Int. J. Theor. Phys.
{\bf 38}, 1113 (1999)].

\bibitem{Gubser1998}
S.S. Gubser, I.R. Klebanov, and A.M. Polyakov,
Phys. Lett. B {\bf 428}, 105 (1998).

\bibitem{Witten}
E. Witten, 
Adv. Theor. Math. Phys. {\bf 2}, 253 (1998).

\bibitem{GubserPRD78}
S.S. Gubser, Phys. Rev. D {\bf 78}, 065034 (2008).

\bibitem{HartnollPRL101}
S.A. Hartnoll, C.P. Herzog, and G.T. Horowitz, Phys. Rev. Lett. {\bf
101}, 031601 (2008).

\bibitem{HartnollJHEP12}
S.A. Hartnoll, C.P. Herzog, and G.T. Horowitz, J. High Energy Phys.
{\bf 12}, 015 (2008).

\bibitem{PanJWCh}
Q.Y. Pan, J.L. Jing, B. Wang, and S.B. Chen,
J. High Energy Phys. {\bf 06}, 087 (2012).

\bibitem{BarclayGregory}
L. Barclay, R. Gregory, S. Kanno, and P. Sutcliffe, J. High Energy
Phys.  {\bf 12}, 029 (2010); arXiv:1009.1991 [hep-th].

\bibitem{Barclay2011}
L. Barclay, J.  High Energy Phys.  {\bf 10}, 044 (2011).

\bibitem{GregoryRev}
R.  Gregory, J. Phys. Conf. Ser.  {\bf 283}, 012016 (2011);
arXiv:1012.1558 [hep-th].

\bibitem{KannoCQG}
S. Kanno, Class. Quant. Grav. {\bf 28}, 127001 (2011);
arXiv:1103.5022 [hep-th].

\bibitem{Ge2011}
X.H. Ge and H.Q. Leng, Prog. Theor. Phys. {\bf 128}, 1211 (2012);
arXiv:1105.4333 [hep-th].

\bibitem{Herzog2010}
C.P.  Herzog, Phys.  Rev.  D {\bf 81}, 126009 (2010);
arXiv:1003.3278 [hep-th].

\bibitem{Gubser-Nellore}
S.S. Gubser and A. Nellore, J. High Energy Phys. {\bf 04}, 008
(2009).

\bibitem{Horowitz-Way}
G.T. Horowitz and B. Way, J. High Energy Phys. {\bf 11}, 011 (2010);
arXiv:1007.3714 [hep-th].

\bibitem{Aprile-Russo}
F. Aprile and J.G.  Russo, Phys.  Rev.  D {\bf 81}, 026009 (2010).

\bibitem{Brihaye}
Y. Brihaye and B. Hartmann, Phys.  Rev.  D {\bf 81}, 126008 (2010).

\bibitem{Liu-Sun}
Y. Liu and Y.W.  Sun, J. High Energy Phys.  {\bf 07}, 099 (2010).

\bibitem{Siani}
M. Siani, J. High Energy Phys.  {\bf 12}, 035 (2010);
arXiv:1010.0700 [hep-th].

\bibitem{PanWangBR}
Q.Y. Pan and B. Wang, arXiv:1101.0222 [hep-th].

\bibitem{LiuWangBTZ}
Y.Q.  Liu, Q.Y.  Pan, and B. Wang, Phys. Lett. B {\bf 702}, 94
(2011).

\bibitem{LPW2012}
Y.Q. Liu, Y. Peng, and B. Wang, arXiv:1202.3586 [hep-th].

\bibitem{GangopadhyayPLB}
S. Gangopadhyay, Phys.  Lett.  B {\bf 724}, 176 (2013);
arXiv:1302.1288 [hep-th].

\bibitem{LiuGongWang}
Y.Q. Liu, Y.G. Gong, and B. Wang, J. High Energy Phys. {\bf 02}, 116
(2016); arXiv:1505.03603 [hep-ph].

\bibitem{YaoJing}
W.P. Yao and J.L. Jing, J. High Energy Phys. {\bf 05}, 101 (2013);
J. High Energy Phys. {\bf 05}, 058 (2014); Nucl. Phys. B {\bf 889},
109 (2014); Phys. Lett. B {\bf 759}, 533 (2016).

\bibitem{EmparanTanabe}
R. Emparan and K. Tanabe, J. High Energy Phys. {\bf 01}, 145 (2014);
arXiv:1312.1108 [hep-th].

\bibitem{DeyJHEP2014}
A. Dey, S. Mahapatra, and T. Sarkar, J. High Energy Phys. {\bf 06},
147 (2014); arXiv:1404.2190 [hep-th].

\bibitem{NakoniecznyRogatko}
L. Nakonieczny and M. Rogatko, Phys. Rev. D {\bf 90}, 106004 (2014);
arXiv:1411.0798 [hep-th].

\bibitem{MomeniPLB2015}
D. Momeni, H. Gholizade, M. Raza, and R. Myrzakulov, Phys. Lett. B
{\bf 747}, 417 (2015); arXiv:1503.02896 [hep-th].

\bibitem{Ghorai2016}
D. Ghorai and S. Gangopadhyay, Eur. Phys. J. C {\bf 76}, 146 (2016).

\bibitem{JingCQG}
J.L. Jing, L. Jiang, and Q.Y. Pan, Class. Quant. Grav. {\bf 33},
025001 (2016).

\bibitem{SheykhiShakerPLB}
A. Sheykhi and F. Shaker, Phys. Lett. B {\bf 754}, 281 (2016).

\bibitem{Peng2017}
Y. Peng, Q.Y. Pan, and Y.Q. Liu, Nucl. Phys. B {\bf 915}, 69 (2017).

\bibitem{SheykhiShakerIJMPD}
A. Sheykhi and F. Shaker, Int. J. Mod. Phys. D {\bf 26}, 1750050
(2017).

\bibitem{SherkatghanadIJMPD}
Z. Sherkatghanad, B. Mirza, and F.L. Dezaki, Int. J. Mod. Phys. D
{\bf 26}, 1750175 (2017).

\bibitem{YaoJing2018}
W.P. Yao, C.H. Yang, and J.L. Jing, Eur. Phys. J. C {\bf 78}, 353
(2018); arXiv:1805.02328 [gr-qc].

\bibitem{Ghotbabadi2018}
B.B. Ghotbabadi, M.K. Zangeneh, and A. Sheykhi, Eur. Phys. J. C {\bf
78}, 381 (2018).

\bibitem{GhoraiNPB2018}
D. Ghorai and S. Gangopadhyay, Nucl. Phys. B {\bf 933}, 1 (2018).

\bibitem{MohammadiSZ2018}
M. Mohammadi, A. Sheykhi, and M.K. Zangeneh, arXiv:1805.07377
[hep-th].

\bibitem{CaiNieZhang2011}
R.G. Cai, Z.Y. Nie, and H.Q. Zhang, Phys. Rev.  D {\bf 83}, 066013
(2011); arXiv:1012.5559 [hep-th].

\bibitem{AriasLandea}
R.E. Arias and I.S. Landea, J. High Energy Phys. {\bf 01}, 157
(2013); arXiv:1210.6823 [hep-th].

\bibitem{CaiPWave-1}
R.G. Cai, L. Li, and L.F. Li, J. High Energy Phys. {\bf 01}, 032
(2014); arXiv:1309.4877 [hep-th].

\bibitem{CaiPWave-2}
R.G. Cai, L. Li, L.F. Li, and R.Q. Yang, J. High Energy Phys. {\bf
04}, 016 (2014); arXiv:1401.3974 [gr-qc].

\bibitem{CSJHEP2015}
P. Chaturvedi and G. Sengupta, J. High Energy Phys. {\bf 04}, 001
(2015).

\bibitem{NieCai2015}
Z.Y. Nie, R.G. Cai, X. Gao, L. Li, and H. Zeng, Eur. Phys. J. C {\bf
75}, 559 (2015); arXiv:1501.00004 [hep-th].

\bibitem{Wang2016}
Y.Q. Wang and S. Liu, J. High Energy Phys. {\bf 11}, 127 (2016).

\bibitem{Nie2017}
Z.Y. Nie, Q.Y. Pan, H.B. Zeng, and H. Zeng, Eur. Phys. J. C {\bf
77}, 69 (2017); arXiv:1611.07278 [hep-th].

\bibitem{GTWjhep2012}
X.H. Ge, S.F. Tu, and B. Wang, J. High Energy Phys. {\bf 09}, 088
(2012); arXiv:1209.4272 [hep-th].

\bibitem{HartnollRev}
S.A. Hartnoll, 
Class. Quant. Grav. {\bf 26}, 224002 (2009).

\bibitem{HerzogRev}
C.P. Herzog, 
J. Phys. A {\bf 42}, 343001 (2009).

\bibitem{HorowitzRev}
G.T. Horowitz, 
Lect. Notes Phys. {\bf 828} 313, (2011); arXiv:1002.1722 [hep-th].

\bibitem{CaiRev}
R.G. Cai, L. Li, L.F. Li, and R.Q. Yang, 
Sci. China Phys. Mech. Astron. {\bf 58}, 060401 (2015);
arXiv:1502.00437 [hep-th].

\bibitem{Balatsky}
A.V. Balatsky, I. Vekhter, and J.X. Zhu,
Rev. Mod. Phys. {\bf 78}, 373 (2006).

\bibitem{Ishii}
T. Ishii and S.J. Sin,
J. High Energy Phys. {\bf 04}, 128 (2013); arXiv:1211.1798 [hep-th].

\bibitem{ZengZhang}
H.B. Zeng and H.Q. Zhang, Nucl. Phys. B {\bf 897}, 276 (2015);
arXiv:1411.3955 [hep-th].

\bibitem{FangJHEP}
L.Q. Fang, X.M. Kuang, B. Wang, and J.P. Wu, J. High Energy Phys.
{\bf 11}, 134 (2015).

\bibitem{KuangE2016}
X.M. Kuang and E. Papantonopoulos, J. High Energy Phys. {\bf 08},
161 (2016); arXiv:1607.04928 [hep-th].

\bibitem{Siopsis}
G. Siopsis and J. Therrien, J. High Energy Phys.
{\bf 05}, 013 (2010).

\bibitem{SiopsisBF}
G. Siopsis, J. Therrien, and S. Musiri, Class.
Quant. Grav. {\bf 29}, 085007 (2012); arXiv:1011.2938 [hep-th].


\bibitem{marginally stable modes}
R.G. Cai, X. He, H.F. Li, and H.Q. Zhang, Phys. Rev. D {\bf 84},
046001 (2011).

\bibitem{Breitenloher}
P. Breitenloher and D. Z. Freedman, Ann. Phys. {\bf 144}, 249
(1982).

\bibitem{KolyvarisKPSA}
T. Kolyvaris, G. Koutsoumbas, E. Papantonopoulos, and G. Siopsis,
Class. Quant. Grav. {\bf 29}, 205011 (2012); arXiv:1111.0263
[gr-qc].

\bibitem{KolyvarisKPSB}
T. Kolyvaris, G. Koutsoumbas, E. Papantonopoulos, and G. Siopsis, J.
High Energy Phys. {\bf 11}, 133 (2013); arXiv:1308.5280 [hep-th].

\bibitem{Gelfand-Fomin}
I.M. Gelfand and S.V. Fomin, \textit{Calculaus of Variations},
Revised English Edition, Translated and Edited by R.A. Silverman,
Prentice-Hall, Inc. Englewood Cliff, New Jersey (1963).

\bibitem{HFLi}
H.F. Li, J. High Energy Phys. {\bf 07}, 135 (2013); arXiv:1306.3071
[hep-th].

\bibitem{HorowitzPRD78}
G.T. Horowitz and M.M. Roberts, Phys. Rev. D {\bf 78}, 126008
(2008).

\bibitem{Pan-WangGB2010}
Q.Y. Pan, B. Wang, E. Papantonopoulos, J. Oliveria, and A.B. Pavan,
Phys. Rev. D {\bf 81}, 106007 (2010).

\bibitem{GubserPufu}
S.S. Gubser and S.S. Pufu, J. High Energy Phys. {\bf 11}, 033
(2008).














\end{thebibliography}
\end{document}